\documentclass[12pt]{iopart}
\usepackage[dvipdfmx]{graphicx}
\usepackage{color}
\usepackage{braket}
\usepackage{ulem}

\usepackage{caption}
\begin{document}

\title[Quantum annealing with twisted fields]{Quantum annealing with twisted fields}

\author{Takashi Imoto, Yuya Seki, Yuichiro Matsuzaki and, Shiro Kawabata}
\address{Research Center for Emerging Computing Technologies(RCECT), National Institute of Advanced Industrial Science and Technology (AIST),
1-1-1 Umezono, Tsukuba, Ibaraki 305-8568, Japan.
}

\ead{matsuzaki.yuichiro@aist.go.jp}



\vspace{10pt}

\begin{abstract}
Quantum annealing is a promising method for solving combinational optimization problems and performing quantum chemical calculations.
The main sources of errors in quantum annealing are the effects of decoherence and non-adiabatic transition.
We propose a method for suppressing both these effects using inhomogeneous twist operators corresponding to the twist angles of transverse fields applied to qubits.
Furthermore, we adopt variational methods to determine the optimal inhomogeneous twist operator for minimizing the energy of the state after quantum annealing. Our approach is useful for increasing the energy gap and/or making the quantum states robust against decoherence during quantum annealing.
In summary, our results can pave the way to a new approach for realizing practical quantum annealing.
\end{abstract}

%
%
\submitto{\NJP}
%
%
%

\section{Introduction}

Quantum annealing (QA) \cite{apolloni1989quantum, finnila1994quantum,kadowaki1998quantum, farhi2000quantum, farhi2001quantum}, which has attracted considerable attention in recent decades,
was originally used
as a method for solving combinational optimization problems.
It is well known that such problems can be mapped to
finding the ground states of Ising Hamiltonians
\cite{lechner2015quantum, kumar2018quantum, choi2011minor}.
In QA, we prepare a ground state of a Hamiltonian of a transverse field, which is called a driving Hamiltonian; then, we let the system evolve with a time-dependent Hamiltonian to change from the transverse field to the problem Hamiltonian. As long as an adiabatic condition is satisfied and the Hamiltonian contains matrix elements to induce transitions between the initial state and the target state, the ground state of the problem Hamiltonian can be prepared after the dynamics.
By measuring the ground state, we can obtain the solutions of combinatorial optimization problems.

Furthermore, many studies have attempted to use QA for quantum chemical calculations. The Hamiltonian of molecules is written in the second quantized form.
There are methods for transforming the Hamiltonian of the second quantized form into that of qubit forms \cite{bravyi2002fermionic, verstraete2005mapping, seeley2012bravyi, tranter2015b, xia2017electronic}.
In quantum chemistry, high accuracy is required in the calculation of molecular energy,
which is referred to as chemical accuracy.
Knowledge of energy with chemical accuracy is essential for predicting the chemical reactions of molecules \cite{eyring1935activated}.

There are several other applications of QA.
For example, QA has been applied to clustering \cite{kurihara2014quantum, kumar2018quantum}.
Further, a method that uses QA to perform calculations for topological data analysis (TDA) has been reported \cite{berwald2018computing}.
In addition, QA has been investigated for solving the shortest-vector problem, which is a candidate for postquantum cryptography \cite{joseph2021two}.

D-Wave Systems Inc. has realized QA machines composed of thousands of qubits \cite{johnson2011quantum}.
Superconducting flux qubits have been employed in these machines.
Many experimental demonstrations of QA have been performed using these machines, including machine learning, and graph coloring problems
\cite{kudo2018constrained, adachi2015application, hu2019quantum, kudo2020localization}.

Although considerable effort has been devoted toward identifying useful applications of QA, many obstacles are yet to be overcome for using QA to solve practical problems. In particular, there are two major obstacles: decoherence and non-adiabatic transitions \cite{morita2008mathematical}.
To suppress decoherence, we need to implement QA with a shorter time schedule. However, as the annealing time becomes shorter, more non-adiabatic transitions will occur during QA.
This trade-off makes it difficult to use QA for solving practical problems.

When the problem Hamiltonian is of the Ising type with only diagonal terms, we randomly obtain the ground state of the problem Hamiltonian after QA by performing measurements in the computational basis.
 The density matrix after QA is expressed as $\rho=\sum_{j}p_{j}\ket{E_{j}}\bra{E_{j}}$, where $p_{j}$ denotes the population and $\ket{E_{j}}$ denotes an eigenvector of the problem Hamiltonian.
In this case, the probability of obtaining the ground state depends on the population of the ground state.
Thus, if the ground-state population is finite, we can obtain the ground state  by increasing the number of trials.

By contrast, if we choose a Hamiltonian with off-diagonal matrix elements as the problem Hamiltonian, we cannot estimate the energy of the Hamiltonian by performing measurements in the computational basis.
In this case, we need to estimate the energy of the Hamiltonian by performing the Pauli measurements.
We consider the case in which the Hamiltonian is composed of the summation of the  products of the Pauli matrices, such as $H=\sum _i c_i \sigma_i$, where $\sigma_i$ is the Pauli product and $c_i$ is the coefficient. We obtain the expectation value of each term by performing measurements of the Pauli products in the quantum states after QA, and we take the sum of the expectation values of all the terms.
The energy of the Hamiltonian
for a density matrix $\rho$
is given by $\braket{H}={\rm{Tr}}[H\rho]=\sum_{n}p_{n}E_{n}$, and we obtain  the expectation values of the Hamiltonian $\braket{H}$ by performing
a large number of measurements.
Hence, if there is a population in the excited state, the energy measured experimentally is different from the ground-state energy.
Therefore, we need to prepare a density matrix close to the exact ground state in order to determine the ground-state energy with high accuracy.

In this paper, we propose a method for suppressing both decoherence and non-adiabatic transitions by
using inhomogeneous twist operators that change the angles of the transverse fields during QA.
We define inhomogeneous twist operators that rotate the direction of the transverse fields of the driving Hamiltonian, and we also define the twist parameters that correspond to the rotation angle at each qubit.
Further, we apply these operators to the driving Hamiltonian.

By using this twisted driving Hamiltonian, we can implement QA for the given twist parameters and measure the energy of the state
after QA.
To minimize the energy, we update the parameters by
using, e.g., gradient descent methods, and we perform QA again with different twist parameters. By repeating these processes, we can obtain the ground-state energy, which is lower than that obtained using conventional QA.
 Through numerical simulations, we demonstrate that our approach suppresses the effect of decoherence and non-adiabatic transitions in QA for some problem Hamiltonians.

The remainder of this paper is organized as follows.
Sections \ref{sec:qa} and \ref{sec:gradient_descent} review QA and gradient descent, respectively.
Section \ref{sec:algo} introduces our scheme with twist operators.
Section \ref{sec:nume_result} describes numerical simulations conducted to evaluate the performance of our scheme and shows that, for some problem Hamiltonians, the ground-state energy obtained using our scheme is more accurate than that obtained using the conventional scheme. Finally, Section \ref{sec:conclusion} concludes the paper.

\section{Quantum Annealing}\label{sec:qa}

Here, we review QA for the ground-state search \cite{kadowaki1998quantum}.
We choose the driving Hamiltonian $H_{D}$ as the transverse field (i.e., $H_{D}=-\sum_{i=1}^{N}\hat{\sigma}_{i}^{x}$).
The total Hamiltonian is described as follows:
\begin{equation}
    H(t)=\Bigl(1-\frac{t}{T}\Bigr) H_{D}+\frac{t}{T}H_{P},
\end{equation}
where $T$ is the annealing time, $H_{D}$ is the driving Hamiltonian, and $H_{P}$ is the problem Hamiltonian.
We prepare the ground state of the driving Hamiltonian.
The driving Hamiltonian is adiabatically changed into the problem Hamiltonian.
If the dynamics is adiabatic, the adiabatic theorem guarantees that
we can obtain the ground state of the problem Hamiltonian.

The accuracy of QA is degraded by many types of noise.
The relevant types are environmental decoherence and non-adiabatic transitions \cite{morita2008mathematical, messiah1961quantum, messiah1962quantum, roland2005noise, aaberg2005quantum, albash2015decoherence, childs2001robustness, sarandy2005adiabatic}.
For the suppression of non-adiabatic transitions, it is necessary to implement QA with a longer time schedule. However, as the annealing time becomes longer, the decoherence effects become more significant during QA.
Owing to this trade-off, it is not straightforward to solve practical problems using QA.

Several methods for suppressing decoherence and non-adiabatic transitions have been investigated.
Susa et al. proposed a way to accelerate the annealing process using an inhomogeneous driving Hamiltonian for a specific case \cite{susa2018exponential, susa2018quantum}.
It is known that "non-stoquastic" Hamiltonians with negative off-diagonal matrix elements improve the performance of QA for some problem Hamiltonians \cite{seki2012quantum, seki2015quantum}.
Direct estimation of the energy gap between the ground state and the first excited state by QA has been proposed \cite{matsuzaki2021direct}, and
this method is robust against non-adiabatic transitions.
In addition, considerable effort has been devoted toward suppressing the effect of environmental noise.
Error correction of QA has been investigated to suppress decoherence \cite{pudenz2014error}.
In addition, the idea of using a decoherence-free subspace for QA has been proposed \cite{albash2015decoherence, suzuki2020proposal}.
Spin lock techniques can be adopted to use long-lived qubits for QA \cite{chen2011experimental, nakahara2013lectures, matsuzaki2020quantum}.
Furthermore, several methods using non-adiabatic transitions and quenching for efficient QA have been studied \cite{crosson2014different, goto2020excited, hormozi2017nonstoquastic, muthukrishnan2016tunneling, brady2017necessary, somma2012quantum, das2008colloquium}.
Other approaches have also been proposed to suppress non-adiabatic transitions and decoherence by using variational methods \cite{susa2021variational, matsuura2021variationally, 2109.13043}.
It is worth noting that such variational methods have been adopted to find a ground state of the Hamiltonian by using variational algorithms with near-term intermediate-scale quantum (NISQ) devices \cite{peruzzo2014variational, mcclean2016theory}.

\section{Gradient descent}\label{sec:gradient_descent}

Here, we review the basic gradient descent method.
The gradient descent method is a method that searches for the lowest value of the cost function by the gradient.
It consists of four steps.
First, we determine the learning rate
that indicates by how much we can change the parameters when we update them.
Second, we derive the gradient of the cost function.
When we cannot obtain the gradient of the cost function analytically, we need to use numerical differentiation.
Third, we update the parameters using the gradient and learning rate as follows:
\begin{eqnarray}
    a^{(1)}=a^{(0)}-\alpha\frac{\partial f(a)}{\partial a}|_{a=a^{(0)}},
\end{eqnarray}
where $f$ is the cost function, $\alpha$ is the learning rate, and $a^{(0)}$ and $a^{(1)}$ are the initial and updated parameters, respectively.
Fourth,
we repeat the second and third steps
until the cost function converges to a specific value.

\section{Our variational twisting scheme}\label{sec:algo}
Here, we introduce our scheme for using twist operators with QA.
By deforming the driving Hamiltonian with an inhomogeneous twist operator, we aim to obtain the ground-state energy with higher accuracy than that obtained using conventional QA.
The inhomogeneous twist operator consists of several parameters, which we refer to as twist parameters.
To find the optimal twist parameters for an efficient ground-state search, we use the so-called variational methods, where we adaptively update the parameters according to the measurement results after QA.

\subsection{Inhomogeneous twist operator}

Let us consider the problem Hamiltonian composed of $L$ qubits.
Let $\sigma_{j}^{x}$,
 $\sigma_{j}^{y}$, and $\sigma_{j}^{z}$ denote
 the standard Pauli matrices at the $j$-th site. Then, the inhomogeneous twist operator is defined by
\begin{eqnarray}
    U_{twist}(\theta_{1},\cdots,\theta_{L}):=\prod_{j=1}^{L}\exp[i\theta_{j}\sigma_{j}^{y}],
\end{eqnarray}
where $L$ is the number of qubits and $\{\theta_{j}\}_{j=1}^L$ are the twist parameters.
We define a single rotational operator of the y-axis at the $j$-th site as
\begin{eqnarray}
    U_{j}^{y}:=\exp[i\theta_{j}\sigma_{j}^{y}].
\end{eqnarray}
In addition, we define the inhomogeneous twist operator as
\begin{eqnarray}
    U_{twist}(\theta_{1},\cdots,\theta_{L}):&=\prod_{j=1}^{L}U_{j}^{y}\nonumber\\
    &=\exp[i\sum_{j=1}^{L}\theta_{j}\sigma_{j}^{y}].
\end{eqnarray}
Further, we deform the driving Hamiltonian with the inhomogeneous twist operator.
The key idea of our scheme is to use the deformed driving Hamiltonian for QA as follows:
\begin{equation}
    H^{(twist)}(t, \theta_{1}, \cdots, \theta_{L})=\Bigl(1-\frac{t}{T}\Bigr) U^{-1}_{twist}(\theta_{1},\cdots,\theta_{L})H_{D}U_{twist}(\theta_{1},\cdots,\theta_{L})+\frac{t}{T}H_{P}. 
    \label{eq:twist_ann_ham}
\end{equation}
The energy spectrum of this total Hamiltonian is changed by the twist parameters.

\subsection{Variational QA with gradient descent}

\begin{figure}[htbp]
\begin{center}
\includegraphics[width=130mm]{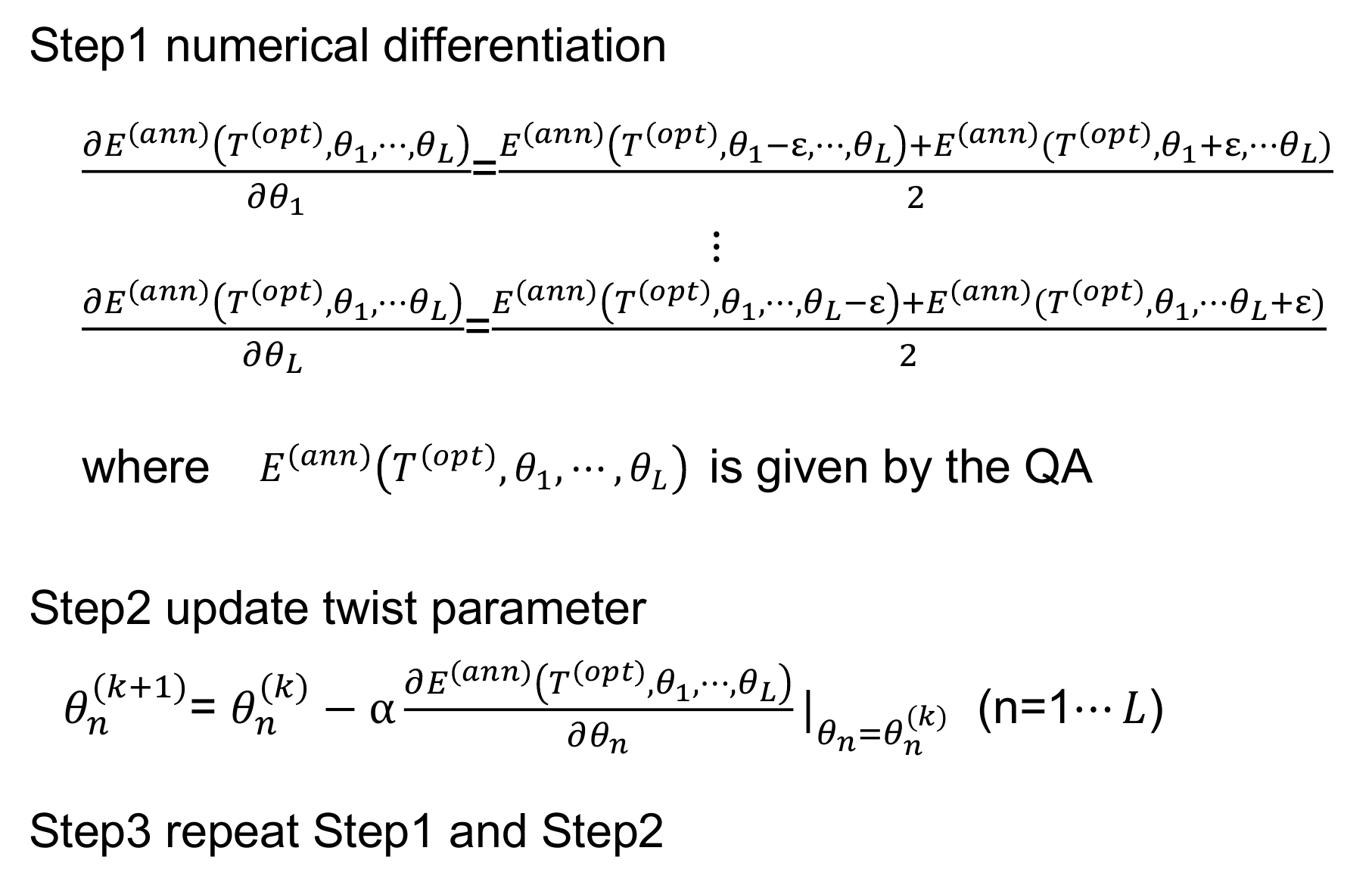}
\caption{Schematic of gradient descent for twisted QA. Here, $\alpha$ denotes the learning rate.}
\label{fig:algo_fig}
\end{center}
\end{figure}

Here, we explain how to apply gradient descent with our scheme.
Let $E^{(ann)}(T,\theta_{1}, \cdots, \theta_{L})$ denote the energy that we measure after QA using the deformed annealing Hamiltonian (\ref{eq:twist_ann_ham}).
Our scheme consists of three steps.
First, to obtain the derivative of $E^{(ann)}(T, \theta_{1}, \cdots, \theta_{L})$ with respect to $\{ \theta_j \}_{j=1}^L$ for a given $T$, we set the inhomogeneous twist operators with some twist parameters, perform QA with the deformed Hamiltonian, and measure the energy of the state after QA.
It is worth noting that we cannot analytically obtain the derivative of $E^{(ann)}(T, \theta_{1}, \cdots, \theta_{L})$; hence, we use numerical differentiation.
Second, we update the twist parameters on the basis of the results of step 1.
Third, we repeat steps 1 and 2 until the energy converges to a finite value.
The entire procedure is summarized in Figure \ref{fig:algo_fig}.

 It is worth noting that by performing the scheme shown in Figure \ref{fig:algo_fig} for several values of the annealing time, we can find the optimized annealing time that minimizes $E^{(ann)}$. This optimized annealing time is denoted by $T^{(opt)}$.

\section{Numerical Results}\label{sec:nume_result}
This section describes numerical simulations conducted to evaluate the performance of our scheme. In particular, to account for decoherence, we employ the Lindblad master equation.
For the problem Hamiltonians, we consider two examples: a hydrogen molecule and a deformed spin star model.

\subsection{Lindblad master equation}

In this subsection, we introduce the Lindblad master equation to consider the decoherence during QA.
The Lindblad master equation that we use in this paper is given by
\begin{eqnarray}
    \frac{d\rho(t)}{dt}=-i[H(t), \rho(t)]+\sum_{n}\gamma[\sigma^{(k)}_{n}\rho(t)\sigma^{(k)}_{n}-\rho(t)],
\end{eqnarray}
where $\sigma^{(k)}_{j}(k=x,y,z)$ denote the Lindblad operators acting at site $j$, $\gamma$ denotes the decoherence rate, and $\rho(t)$ is the density matrix of the quantum state at time $t$.
We solve the Lindblad master equation using QuTiP \cite{johansson2012qutip, johansson184nation}.
Throughout this paper, we choose $\sigma_{j}^{z}$ as the Lindblad operator because this type of noise is considered as the main source of decoherence for the qubits used in QA \cite{puri2017quantum}.

\subsection{Hydrogen molecule}
In this subsection, we discuss the numerical results obtained using a hydrogen molecule as the problem Hamiltonian.
In our numerical simulations, the dynamics of the state strongly depends on the annealing time.

First, we introduce the hydrogen molecule.
The Hamiltonian of the hydrogen molecule is described by the second quantized form.
We map the Hamiltonian with the second quantized form to a spin Hamiltonian by the Jordan--Wigner transformation.
The Hamiltonian of the hydrogen molecule is given by
\begin{eqnarray}
    H&=& h_{0}I+h_{1}\hat{\sigma}^{z}_{0}
    +h_{2}\hat{\sigma}^{z}_{1}
    +h_{3}\hat{\sigma}^{z}_{2}
    +h_{4}\hat{\sigma}^{z}_{3}\nonumber\\
    &+&h_{5}\hat{\sigma}^{z}_{0}\hat{\sigma}^{z}_{1}
    +h_{6}\hat{\sigma}^{z}_{0}\hat{\sigma}^{z}_{2}
    +h_{7}\hat{\sigma}^{z}_{1}\hat{\sigma}^{z}_{2}
    +h_{8}\hat{\sigma}^{z}_{0}\hat{\sigma}^{z}_{3}
    +h_{9}\hat{\sigma}^{z}_{1}\hat{\sigma}^{z}_{3}\nonumber\\
    &+&h_{10}\hat{\sigma}^{z}_{2}\hat{\sigma}^{z}_{3}
    +h_{11}\hat{\sigma}^{y}_{0}\hat{\sigma}^{y}_{1}\hat{\sigma}^{x}_{2}\hat{\sigma}^{x}_{3}
    +h_{12}\hat{\sigma}^{x}_{0}\hat{\sigma}^{y}_{1}\hat{\sigma}^{y}_{2}\hat{\sigma}^{x}_{3}\nonumber\\
    &+&h_{13}\hat{\sigma}^{y}_{0}\hat{\sigma}^{x}_{1}\hat{\sigma}^{x}_{2}\hat{\sigma}^{y}_{3}
    +h_{14}\hat{\sigma}^{x}_{0}\hat{\sigma}^{x}_{1}\hat{\sigma}^{y}_{2}\hat{\sigma}^{y}_{3},\label{eq:hydrogen_hamiltonian}
\end{eqnarray}
where we use the STO-3G basis and Jordan--Wigner transformation.
The coefficients $h_{0}, h_{1}, \cdots, h_{14}$ of the Hamiltonian in Eq. (\ref{eq:hydrogen_hamiltonian}) depend on the interatomic distance.
We obtain these coefficients of the Hamiltonian expressed by the spin in Eq. (\ref{eq:hydrogen_hamiltonian})
using OpenFermion for each  interatomic distance.
We choose the interatomic distance as $0.74$ \AA, and the coefficients of the Hamiltonian with this interatomic distance are listed in Table \ref{tb:hydrogen_coefficient_jw}.
It is known that the ground state of the Hamiltonian of the hydrogen molecule is very close to the separable state using the Hartree--Fock approximation.

\begin{table}[htb]
\centering
  \caption{Coefficients of the hydrogen molecule using the Jordan--Wigner transformation. The unit of these values is GHz}

    \begin{tabular}{|c||c|}
    \hline
      $h_{0}$ & $-0.09706626816762881$ \\
      $h_{1}$ & $0.17141282644776895$ \\
      $h_{2}$ & $0.17141282644776892$ \\
      $h_{3}$ & $-0.22343153690813586$ \\
      $h_{4}$ & $-0.22343153690813589$ \\
      $h_{5}$ & $0.16868898170361213$ \\
      $h_{6}$ & $0.12062523483390428$ \\
      $h_{7}$ & $0.16592785033770355$ \\
      $h_{8}$ & $0.16592785033770355$ \\
      $h_{9}$ & $0.12062523483390428$ \\
      $h_{10}$ & $0.17441287612261597$ \\
      $h_{11}$ & $-0.04530261550379928$ \\
      $h_{12}$ & $0.04530261550379928$ \\
      $h_{13}$ & $0.04530261550379928$ \\
      $h_{14}$ & $-0.04530261550379928$ \\ \hline
    \end{tabular}
  \label{tb:hydrogen_coefficient_jw}
\end{table}

\begin{figure}[htbp]
  \begin{center}
    \begin{tabular}{c}

      \begin{minipage}{0.5\hsize}
        \begin{center}
          \includegraphics[clip, width=7.5cm]{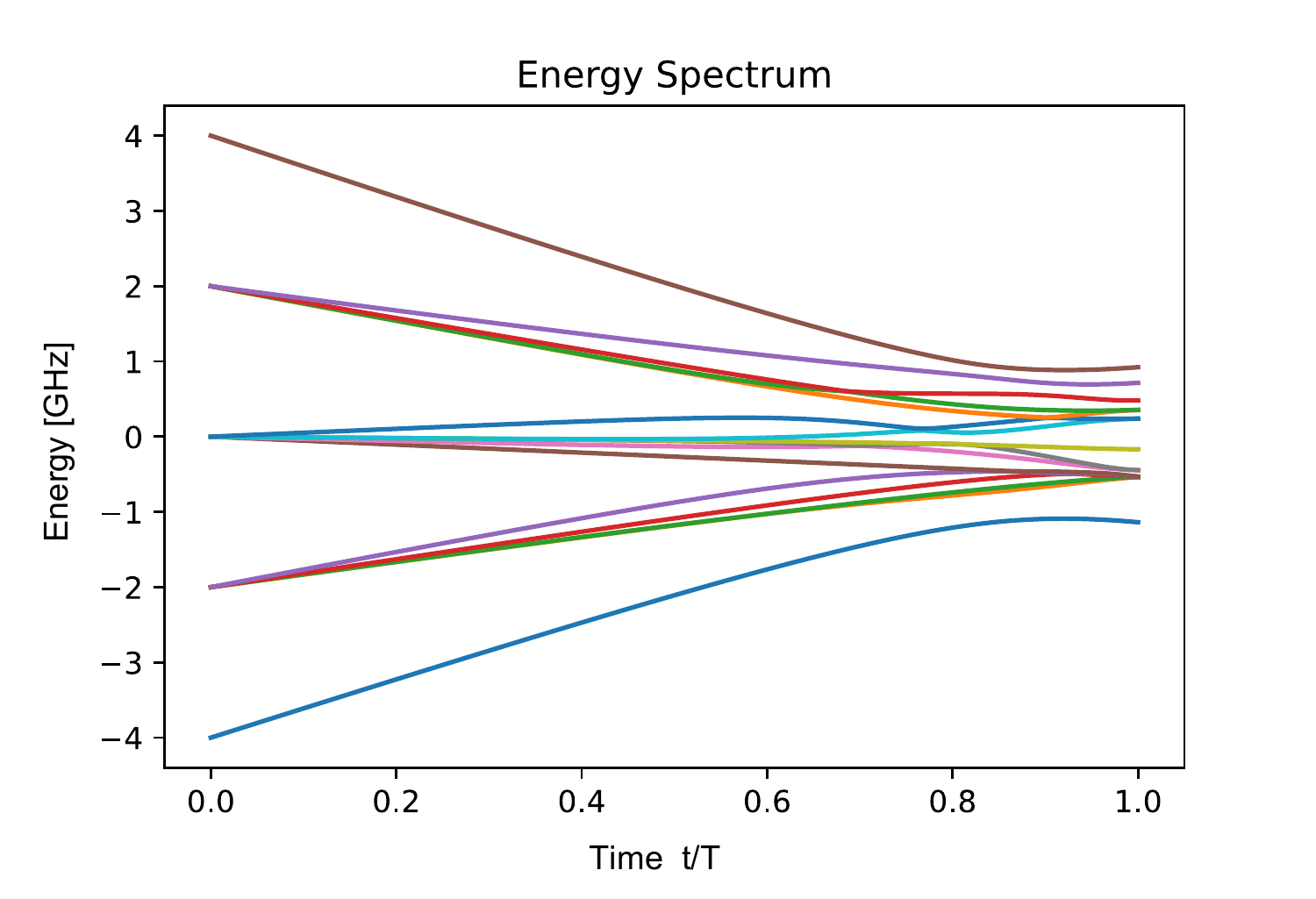}
          \hspace{1.6cm}
        \end{center}
      \end{minipage}

      \begin{minipage}{0.5\hsize}
        \begin{center}
          \includegraphics[clip, width=7.5cm]{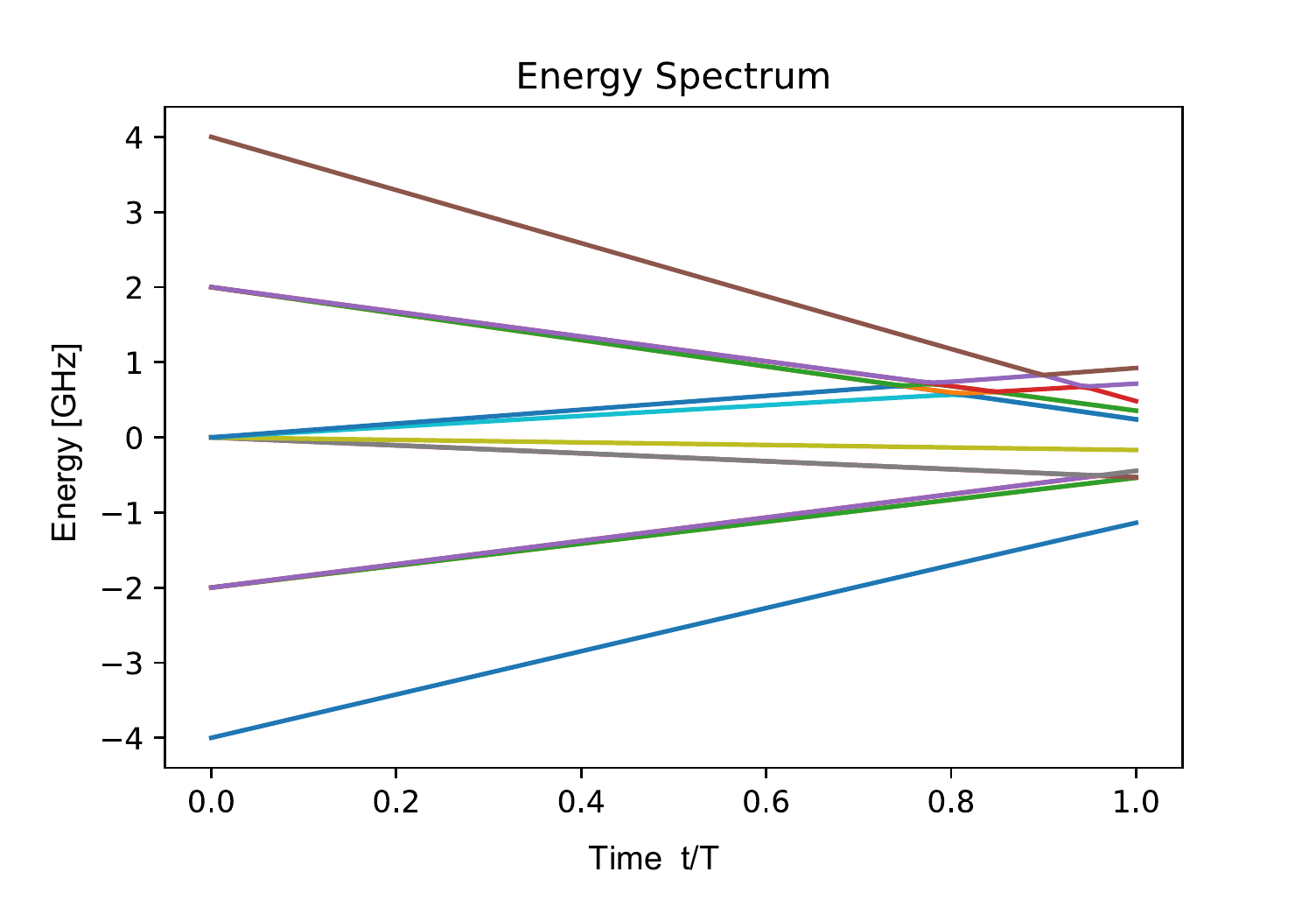}
          \hspace{1.6cm}
        \end{center}
      \end{minipage}

    \end{tabular}
    \caption{Energy spectrum between the driving Hamiltonian and the problem Hamiltonian plotted at each time $t$. The hydrogen molecule is chosen as the problem Hamiltonian. The transverse field (left) and the optimal twisted transverse field (right) are chosen as the driving Hamiltonian.}
    \label{fig:hydroge_energy_spectrum}
  \end{center}
\end{figure}

\begin{figure}[htbp]
  \begin{center}
    \begin{tabular}{c}

      \begin{minipage}{0.5\hsize}
        \begin{center}
          \includegraphics[clip, width=7.5cm]{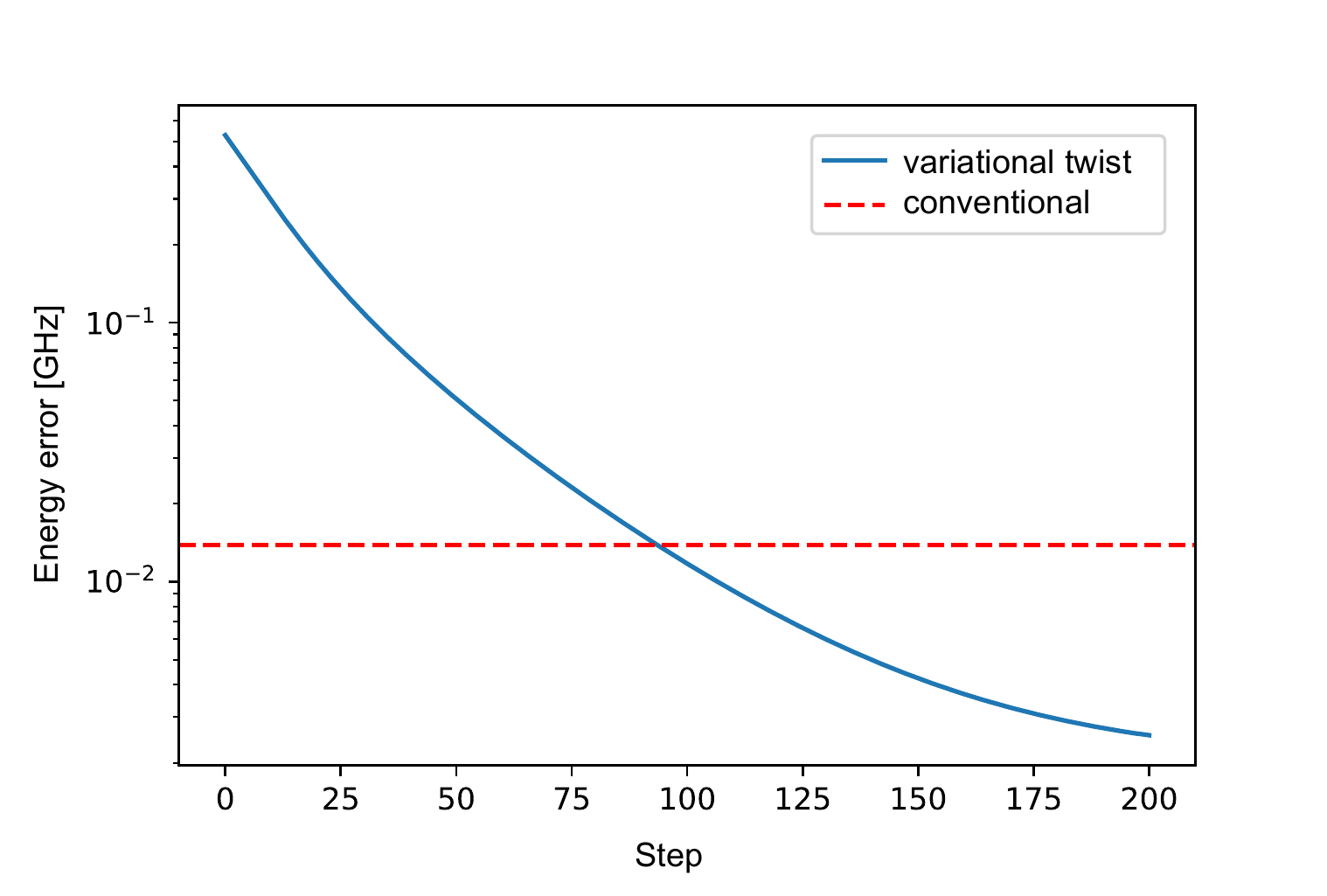}
          (a) Estimation error of the energy against the variational step
          \hspace{1.6cm}
        \end{center}
      \end{minipage}

      \begin{minipage}{0.5\hsize}
        \begin{center}
          \includegraphics[clip, width=7.5cm]{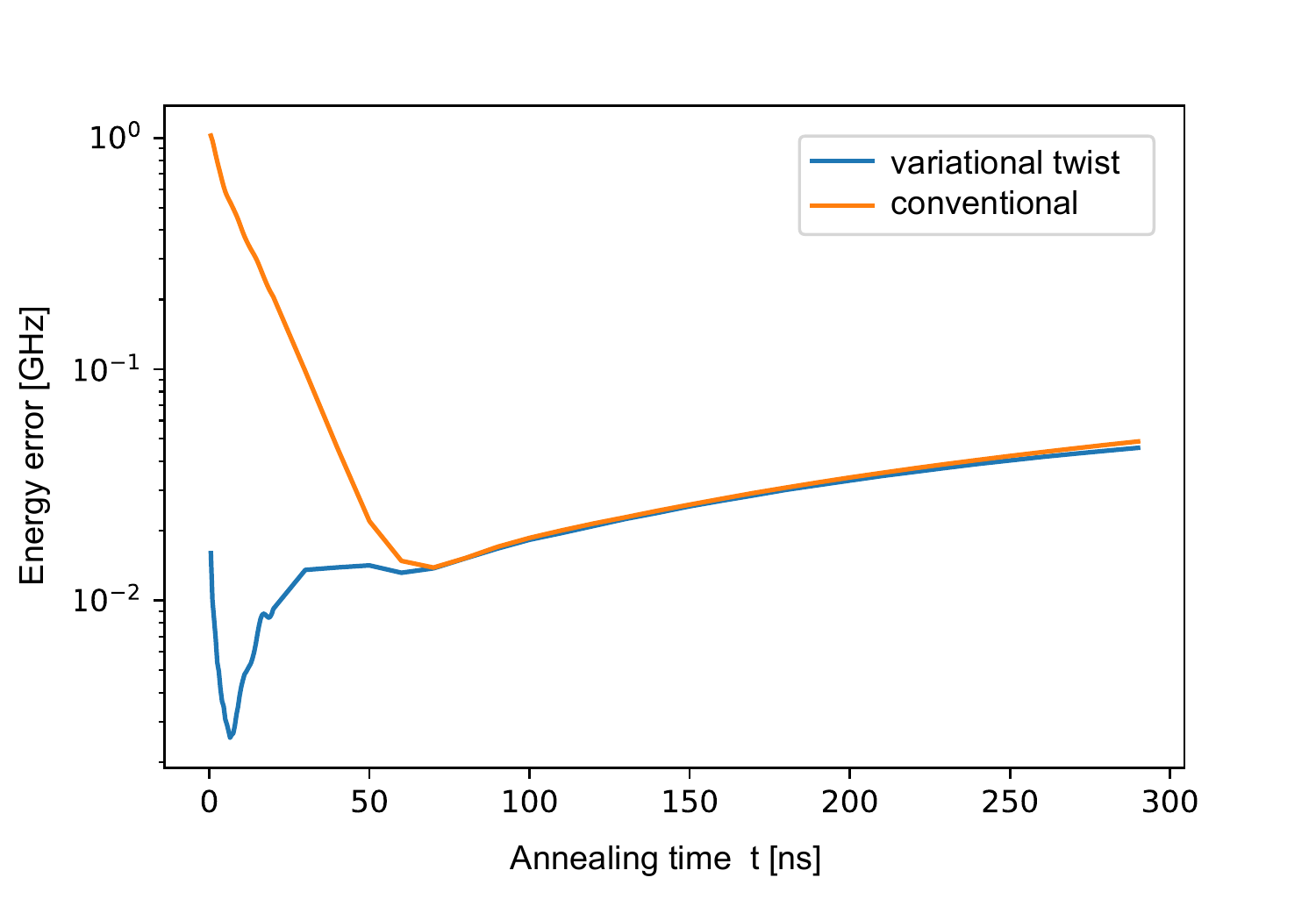}
          (b) Estimation error of the energy against the annealing time $t$
          \hspace{1.6cm}
        \end{center}
        \label{fig:hydroge_error_time}
      \end{minipage}
    \end{tabular}
    \caption{(a) Estimation error of the energy plotted against the variational step on a log scale. The annealing time is chosen to minimize the energy in our scheme. (b) Estimation error of the energy plotted against the annealing time $t$ on a log scale. In (a) and (b), the learning rate $\alpha=0.05$, the decoherence rate $\gamma=10^{-4}$, and the number of steps is $200$.}
    \label{fig:hydroge_error_step_time}
  \end{center}
\end{figure}

In this case, as will be discussed later, the twist parameters tend to be chosen such that the ground states of the driving Hamiltonian and problem Hamiltonian are very close.
We conduct numerical simulations to quantify the performance of our scheme, where we use inhomogeneous twist parameters for the driving Hamiltonian to lower the energy in a variational manner. For comparison, we also perform numerical simulations with conventional QA, where the driving Hamiltonian is chosen as the transverse field.
We set $\gamma=10^{-4}$ and $\alpha=0.05$. Furthermore, the number of steps is $200$.
We plot the energy spectrum of the annealing Hamiltonian
for each scheme in Figure \ref{fig:hydroge_energy_spectrum}.
We observe that the twist operations increase the energy gap between the ground state energy and the first excited state energy.
In this paper, we define an estimation error as the difference between the true ground state energy and the measured energy after QA. If our scheme provides a smaller estimation error than the conventional scheme, it is considered to be more accurate than the conventional scheme.
In Figure \ref{fig:hydroge_error_step_time}, we plot the estimation error against the number of variational steps and annealing time using either our variational scheme with the optimal variational parameters or 
the conventional approach, and we find that the estimation error of our scheme is one order of magnitude smaller than that of the conventional scheme.
In addition, we find that the optimal annealing time for minimizing the energy in our scheme is shorter than that in the conventional approach.

When we set a short annealing time, our variational method tends to choose a driving Hamiltonian whose ground state has a large overlap with the target ground state of the problem Hamiltonian. By contrast, when we set a long annealing time, our method tends to choose a driving Hamiltonian whose ground state is robust against decoherence.

\begin{figure}[htbp]
  \begin{center}
    \begin{tabular}{c}

      \begin{minipage}{0.5\hsize}
        \includegraphics[clip, width=7.5cm]{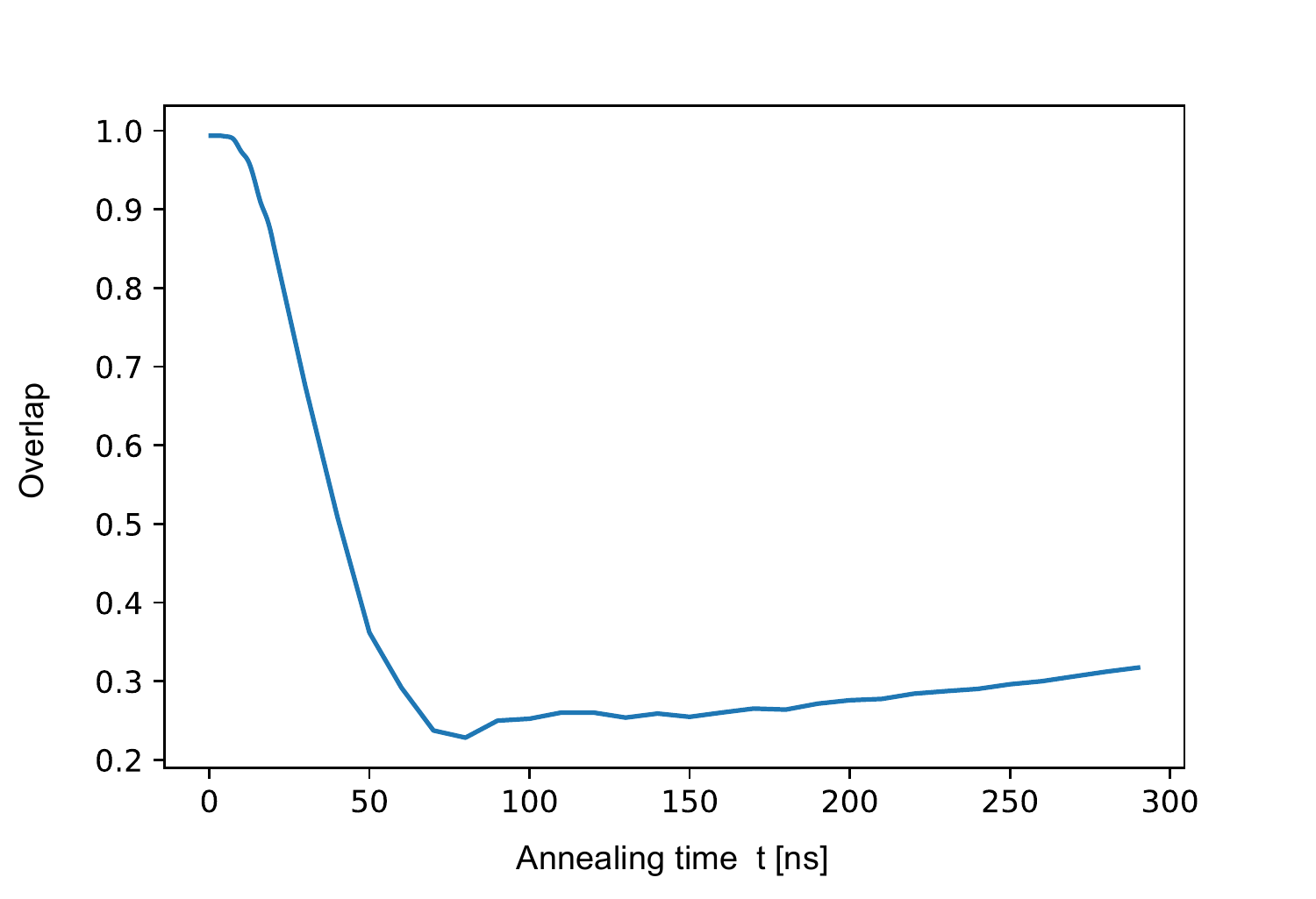}\\
        (a) Overlap between the initial state of twisted QA and the ground state of the problem Hamiltonian
      \end{minipage}

      \begin{minipage}{0.5\hsize}
        \begin{center}
          \includegraphics[clip, width=7.5cm]{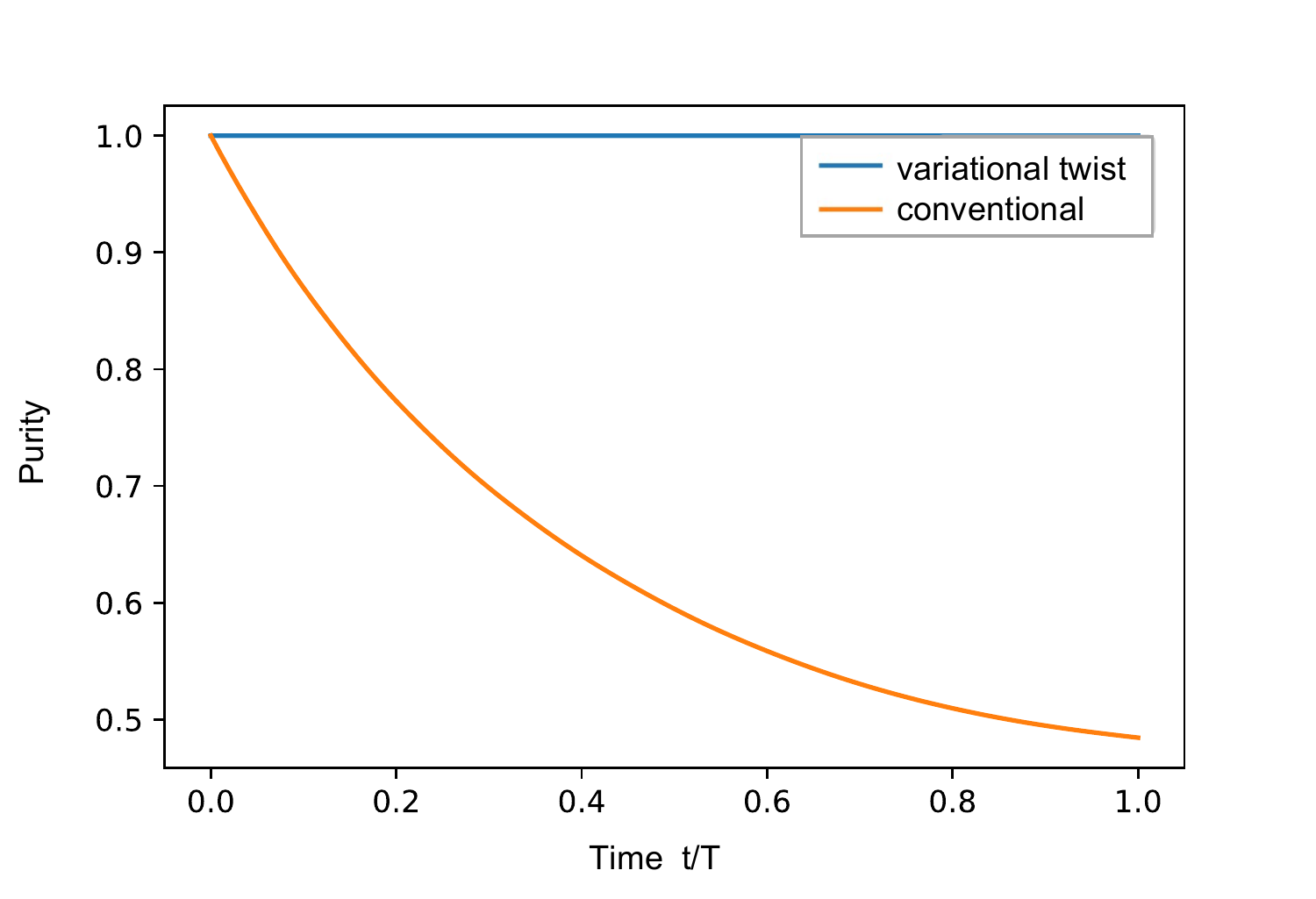}
          (b) Purity against the annealing time
          \hspace{1.6cm}
        \end{center}
      \end{minipage}
    \end{tabular}
  \end{center}
  \caption{(a) Overlap between the initial state of twisted QA and the ground state of the problem Hamiltonian. (b) Purity plotted against the annealing time. The annealing time is chosen to minimize the energy in our scheme. In (a) and (b), the learning rate $\alpha=0.05$, the decoherence rate $\gamma=10^{-4}$, and the number of steps is $200$.}
  \label{fig:overlap_purity_hydro}
\end{figure}

Let $ \mid
\braket{\varphi_{drive}|\varphi_{prob}} \mid $ denote the overlap, where $\ket{\varphi_{drive}}$ denotes the ground state of the driving Hamiltonian and $\ket{\varphi_{prob}}$ denotes the ground state of the problem Hamiltonian.
In Figure \ref{fig:overlap_purity_hydro} (a), we plot the overlap between the initial state of twisted QA and the ground state of the problem Hamiltonian.
It is worth noting that
this overlap can be large only when the ground state of the problem Hamiltonian is close to the product states.
From Figure \ref{fig:overlap_purity_hydro} (a), we find that the overlap becomes especially large for a short annealing time. To investigate the reason for this increase, we use the so-called purity, which is known as a measure for quantifying the effect of decoherence. It is defined by
\begin{eqnarray}
     P=\Tr({\rho^{2}}),
\end{eqnarray}
where $\rho$ is a density matrix.
For a pure state, the purity becomes $1$, whereas it becomes exponentially small against the number of qubits for a completely mixed state.
We plot the purity to quantify the effect of decoherence (see Figure \ref{fig:overlap_purity_hydro} (b)).
In conventional QA, as we increase the annealing time, the purity decreases owing to decoherence. By contrast, in our scheme, the decoherence effect is negligible. This is probably because both the initial state of the driving Hamiltonian in our scheme and the ground state of the problem Hamiltonian are nearly eigenstates of $\hat{\sigma}_z$, which is robust against  $\hat{\sigma}_z$ noise.

\begin{figure}[htbp]
  \begin{center}
    \begin{tabular}{c}

      \begin{minipage}{0.5\hsize}
        \includegraphics[clip, width=7.5cm]{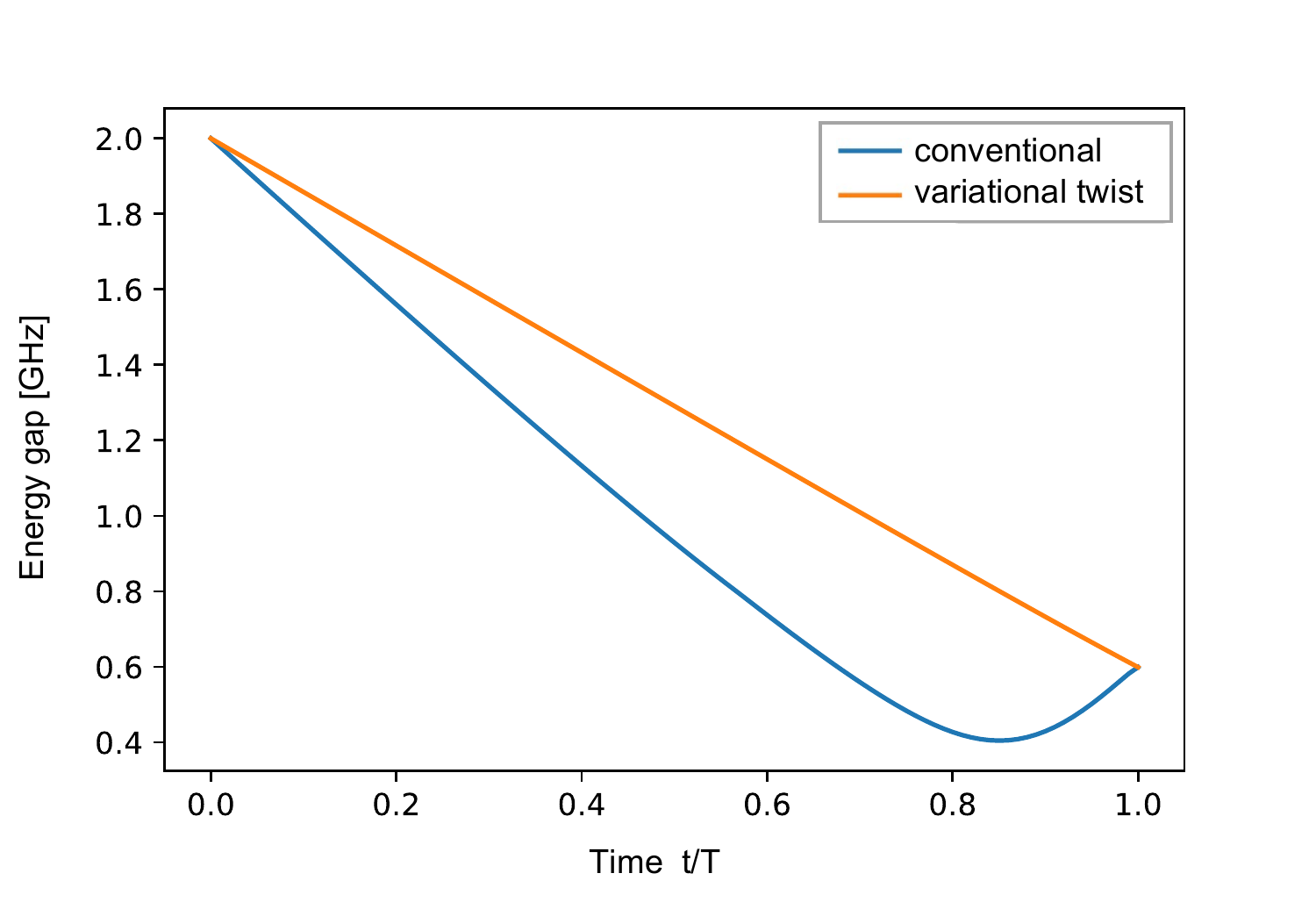}
        (a) Energy gap between the ground state and the first excited state
      \end{minipage}

      \begin{minipage}{0.5\hsize}
        \begin{center}
          \includegraphics[clip, width=7.5cm]{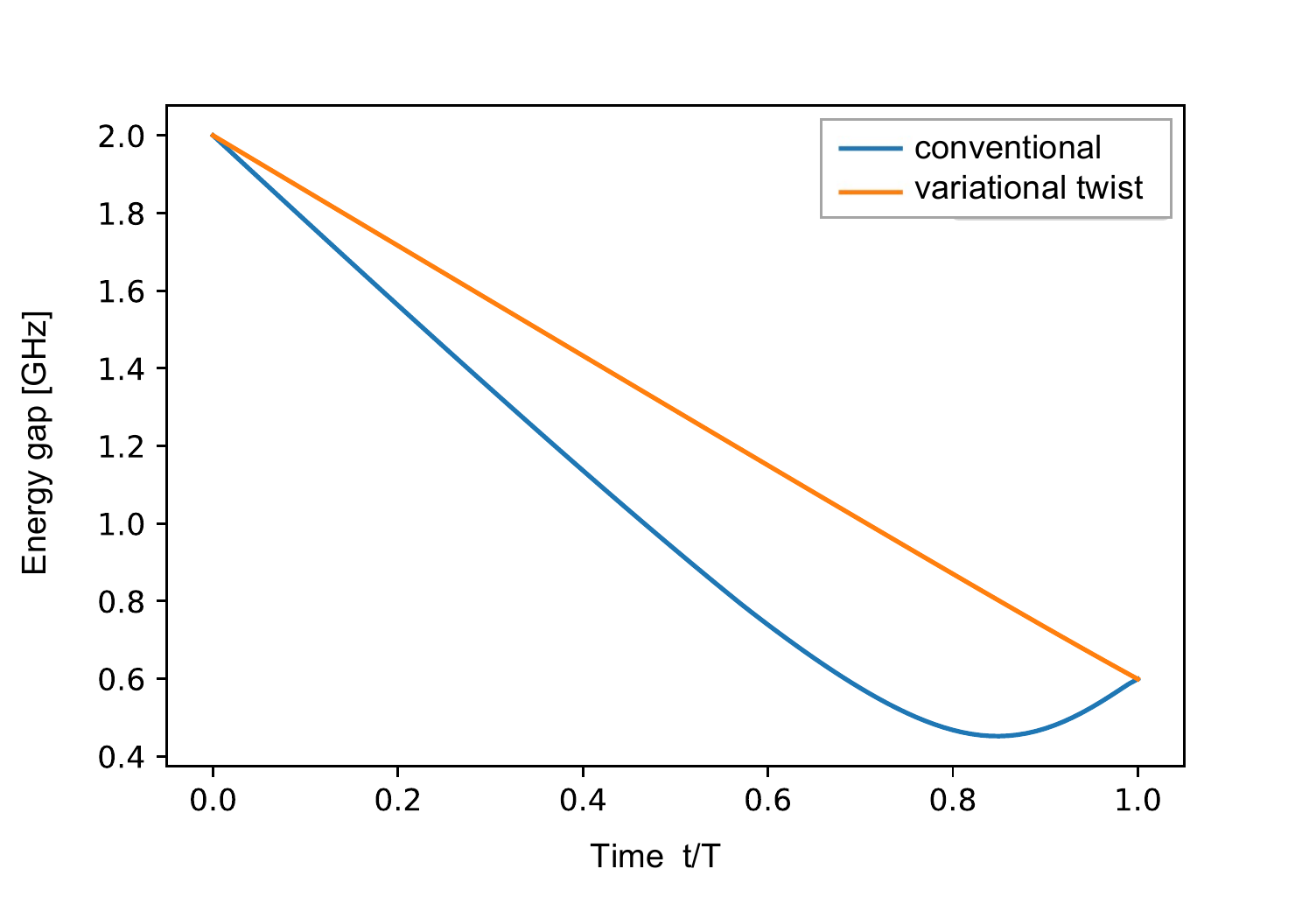}
          (b) Energy gap between the ground state and the second excited state
          \hspace{1.6cm}
        \end{center}
      \end{minipage}
    \end{tabular}
  \end{center}
  \caption{(a) Energy gap between the ground state and the first excited state plotted at each time $t$. (b) Energy gap between the ground state and the second excited state plotted at each time $t$. In (a) and (b), the annealing time is chosen to minimize the energy in our scheme and the conventional scheme.}
  \label{fig:energy_gap_hydro}
\end{figure}

\begin{figure}[htbp]
  \begin{center}
    \begin{tabular}{c}

      \begin{minipage}{0.5\hsize}
        \includegraphics[clip, width=7.5cm]{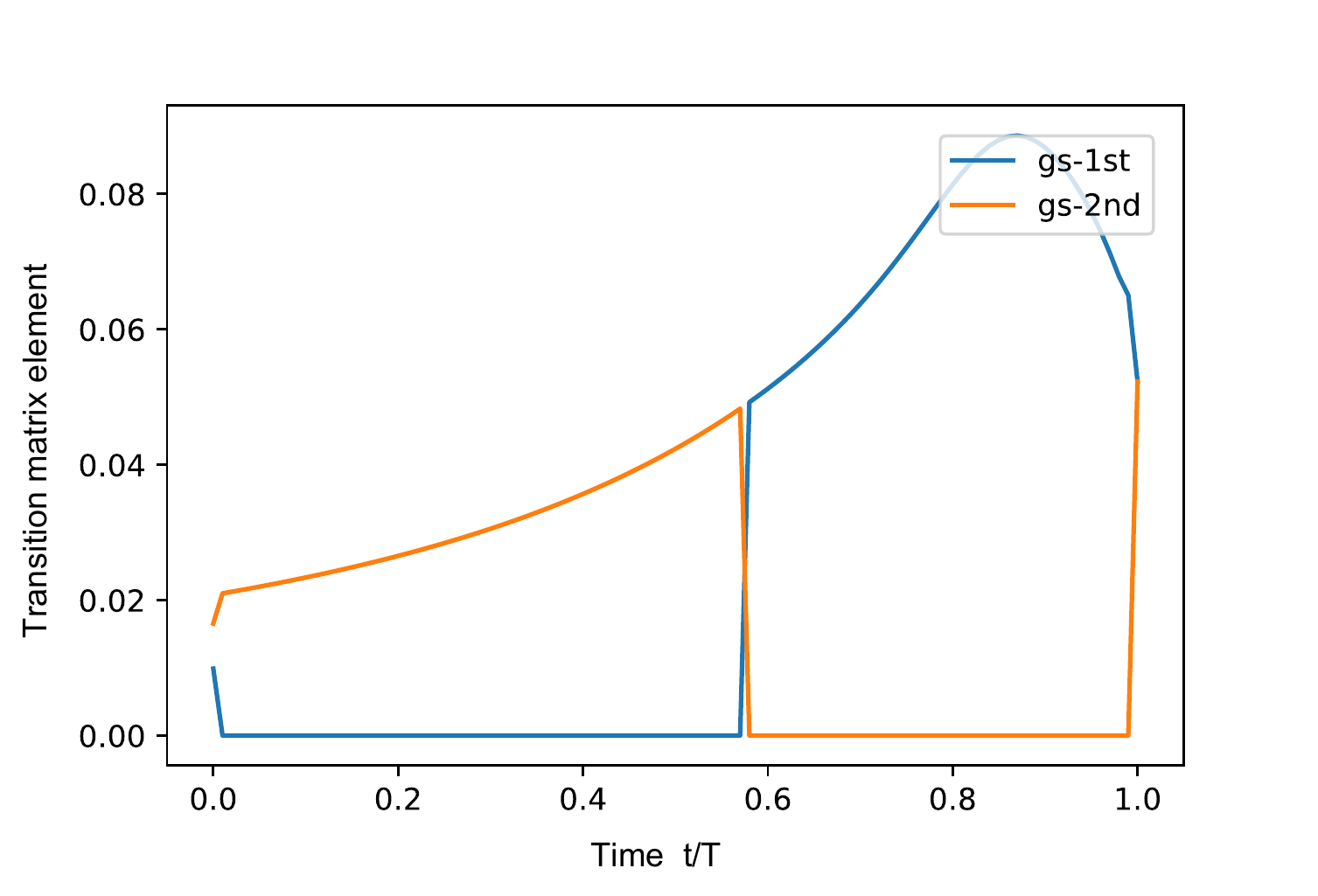}\\
        (a) Element of the transition matrix of the Hamiltonian in the conventional case
      \end{minipage}

      \begin{minipage}{0.5\hsize}
        \begin{center}
          \includegraphics[clip, width=7.5cm]{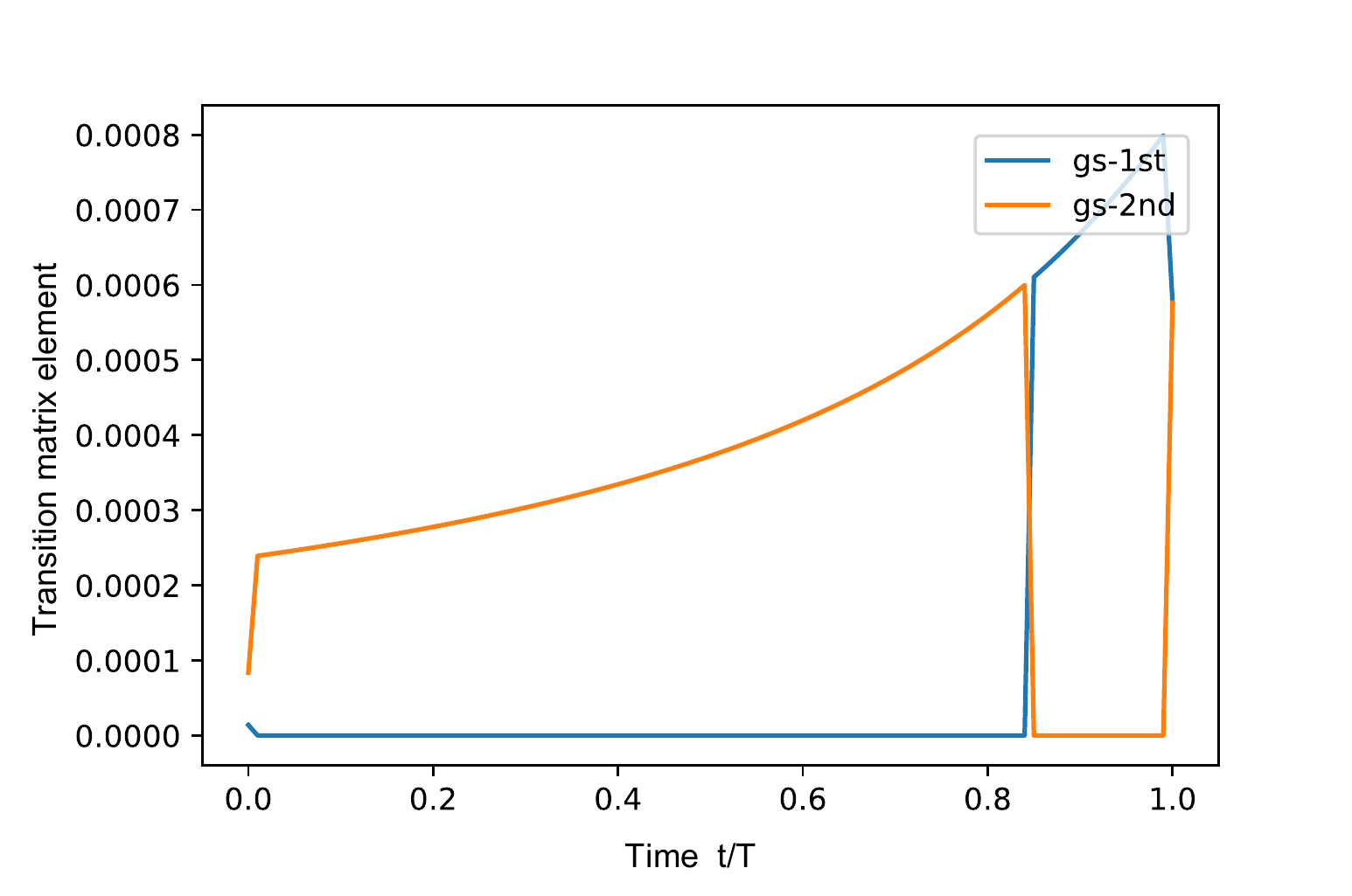}\\
          (b) Element of the transition matrix of the Hamiltonian in the twisted case
          \hspace{1.6cm}
        \end{center}
      \end{minipage}
    \end{tabular}
  \end{center}
  \caption{(a) Element of the transition matrix of the derivative of the Hamiltonian in the conventional scheme. (b) Element of the transition matrix of the derivative of the Hamiltonian in our scheme. In (a) and (b), the annealing time is chosen to minimize the energy in the conventional scheme and our scheme.}
  \label{fig:transition_matrix_hydro}
\end{figure}

Furthermore, we analyzed a so-called adiabatic condition.
If the following quantity $A_j$ is much smaller than $1$ for all $j$, the adiabatic condition is satisfied \cite{morita2008mathematical, messiah1961quantum, messiah1962quantum, jansen2007bounds}:
\begin{eqnarray}
A_j=     \frac{|\bra{E_{j}}\frac{\partial{H}}{\partial{t}}\ket{E_{0}}|}{(E_{j}-E_{0})^{2}},
\end{eqnarray}
where the numerator denotes the transition matrix elements of the derivative of the Hamiltonian
from the ground state to the $j$-th excited state and the denominator denotes the energy gap between the ground state and the $j$-th excited state.
We plot the energy gap between the ground state and the first excited state(see Figure \ref{fig:energy_gap_hydro} (a)).
We can see that our method increases the energy gap between the ground state and the first excited state.
Similarly,  the energy gap between the ground state and the second excited state is increased in our scheme, as shown in Figure \ref{fig:energy_gap_hydro} (b).
Furthermore, we plot the transition matrix that corresponds to the numerator of $A_j$ for conventional QA and
our method in Figure \ref{fig:transition_matrix_hydro} (a) and (b), respectively.
We find that the transition matrix in our scheme is around two orders of magnitude smaller than that in the conventional one.

Therefore, in the case of the hydrogen molecule, our method is advantageous in all aspects, namely the effects of decoherence, the energy gap, and the transition matrix of the derivative of the annealing Hamiltonian.

\subsection{Deformed spin star model}

In this subsection, we consider a deformed spin star model as the problem Hamiltonian.
In contrast to the case of the hydrogen molecule, the ground state of the deformed spin star model is highly entangled, and the overlap between the initial ground state of the twisted driving Hamiltonian and the ground state of the problem Hamiltonian cannot be large regardless of the twist parameters.
The deformed spin star model Hamiltonian is given by
\begin{eqnarray}
    H=\omega\hat{\sigma}_{0}^{z}+\omega_{1}\hat{J}^{z}+J(\hat{\sigma}_{0}^{+}\hat{J}^{-}+\hat{\sigma}_{0}^{-}\hat{J}^{+}),
\end{eqnarray}
where $\hat{J}^{+}\equiv\sum_{j=1}^{N}e^{2\pi\frac{j}{N}}\sigma_{j}^{+}$ and $\hat{J}^{-}\equiv\sum_{j=1}^{N}e^{-2\pi\frac{j}{N}}\sigma_{j}^{-}$.
This model has been studied to represent a hybrid system composed of a superconducting flux qubit and nitrogen-vacancy centers in a diamond lattice \cite{marcos2010coupling,twamley2010superconducting,zhu2011coherent,zhu2014observation,matsuzaki2015improving,cai2015analysis}.
It is known that the ground state of the deformed spin star model with $\omega=\omega_1$ is
\begin{eqnarray}
    \ket{W_{\theta}}=\frac{1}{L}\sum_{j=1}^{L}\sigma_{j}^{+}e^{ij\theta}\ket{\downarrow\downarrow\downarrow\cdots\downarrow}.
\end{eqnarray}
As this state is highly entangled, an overlap with a product state created by the driving Hamiltonian in our scheme cannot be large.

\begin{figure}[htbp]
  \begin{center}
    \begin{tabular}{c}

      \begin{minipage}{0.5\hsize}
        \begin{center}
          \includegraphics[clip, width=7.5cm]{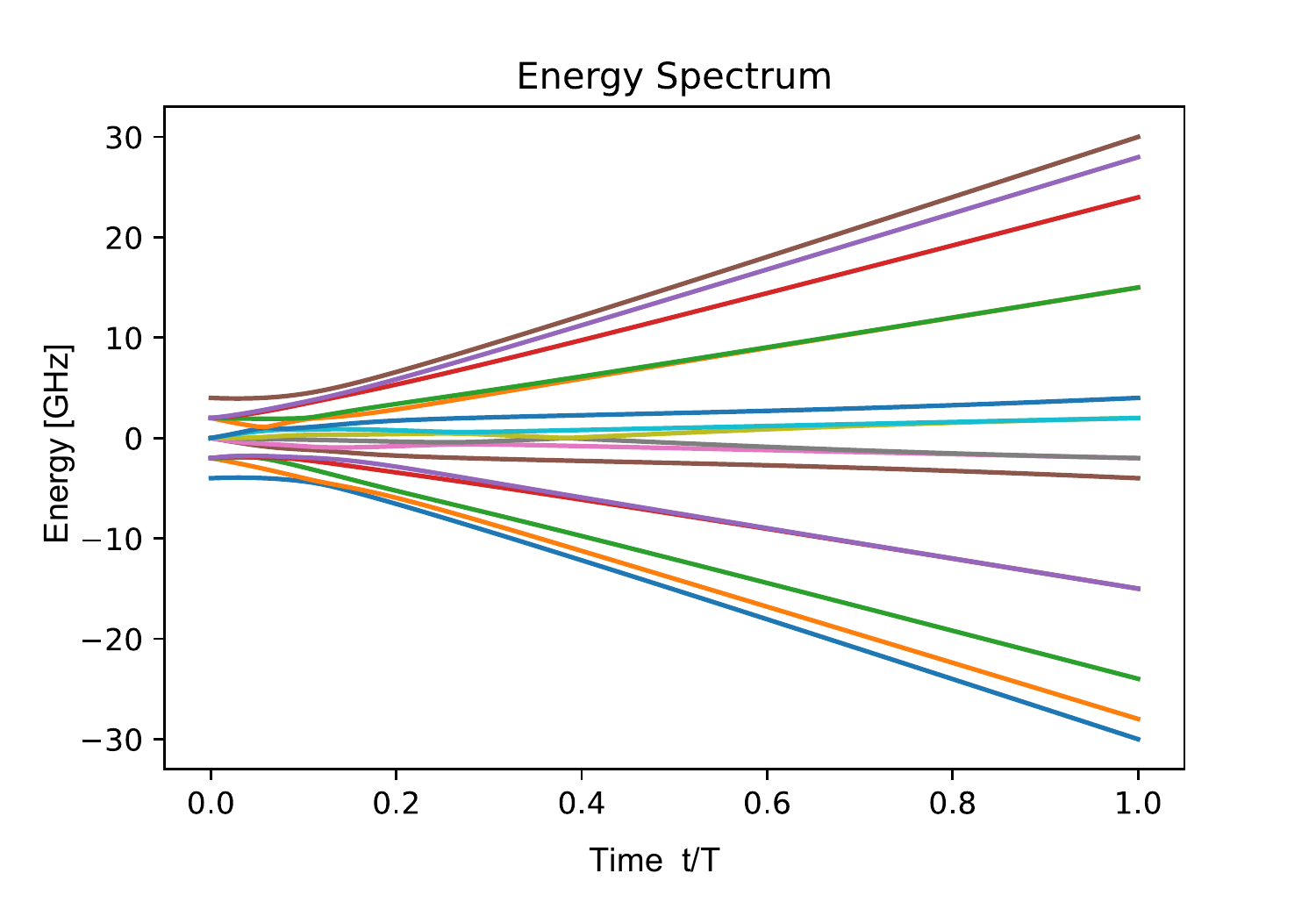}
          \hspace{1.6cm}
        \end{center}
      \end{minipage}


      \begin{minipage}{0.5\hsize}
        \begin{center}
          \includegraphics[clip, width=7.5cm]{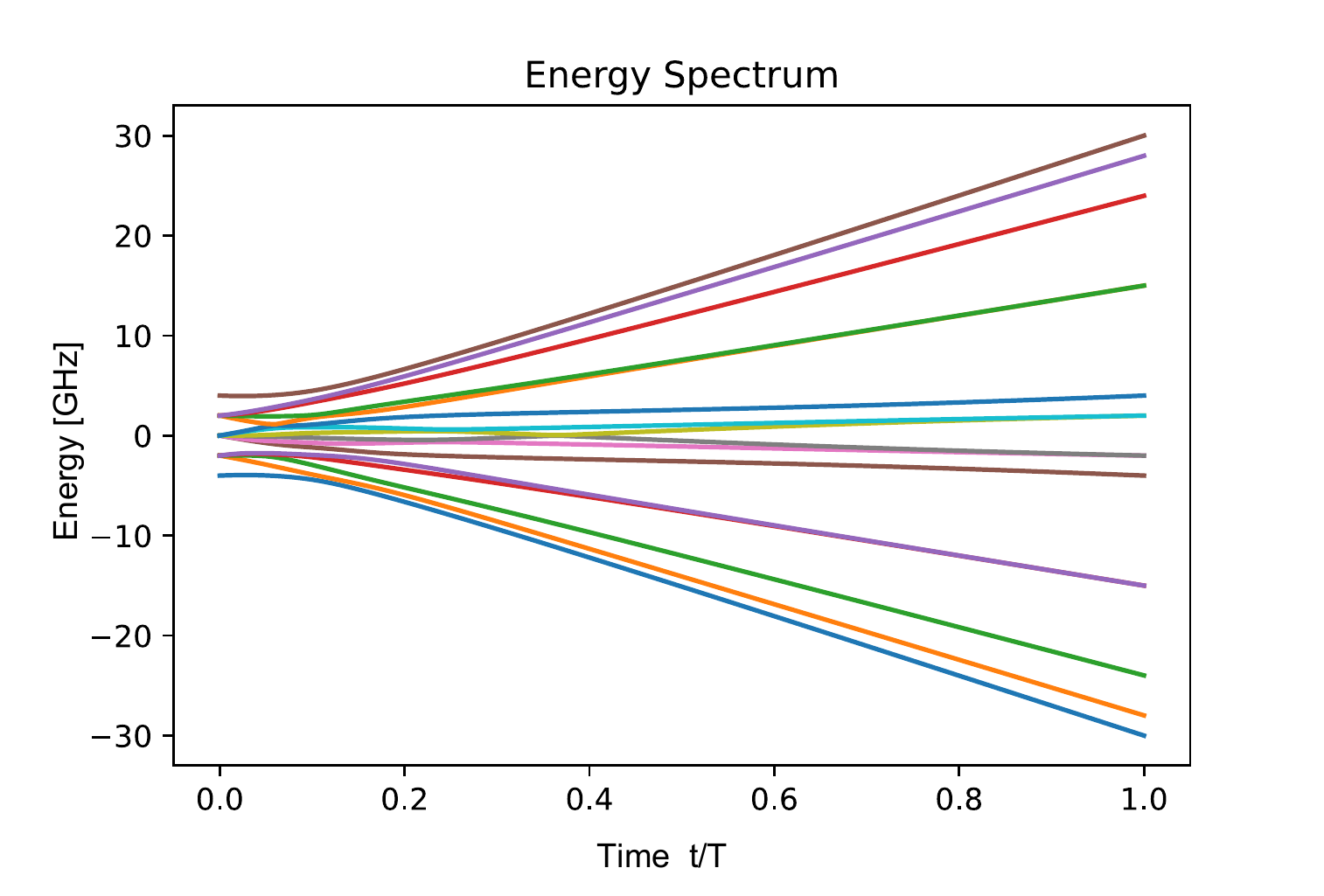}
          \hspace{1.6cm}
        \end{center}
      \end{minipage}

    \end{tabular}
    \caption{Energy spectrum during QA plotted at each time $t$.
    The deformed spin star model is chosen as the problem Hamiltonian. The transverse field (left) and the optimal twisted transverse field (right) is chosen as the driving Hamiltonian.}
    \label{fig:deformed_spin_star_energy_spectrum}
  \end{center}
\end{figure}

\begin{figure}[htbp]
  \begin{center}
    \begin{tabular}{c}

      \begin{minipage}{0.5\hsize}
        \begin{center}
          \includegraphics[clip, width=7.5cm]{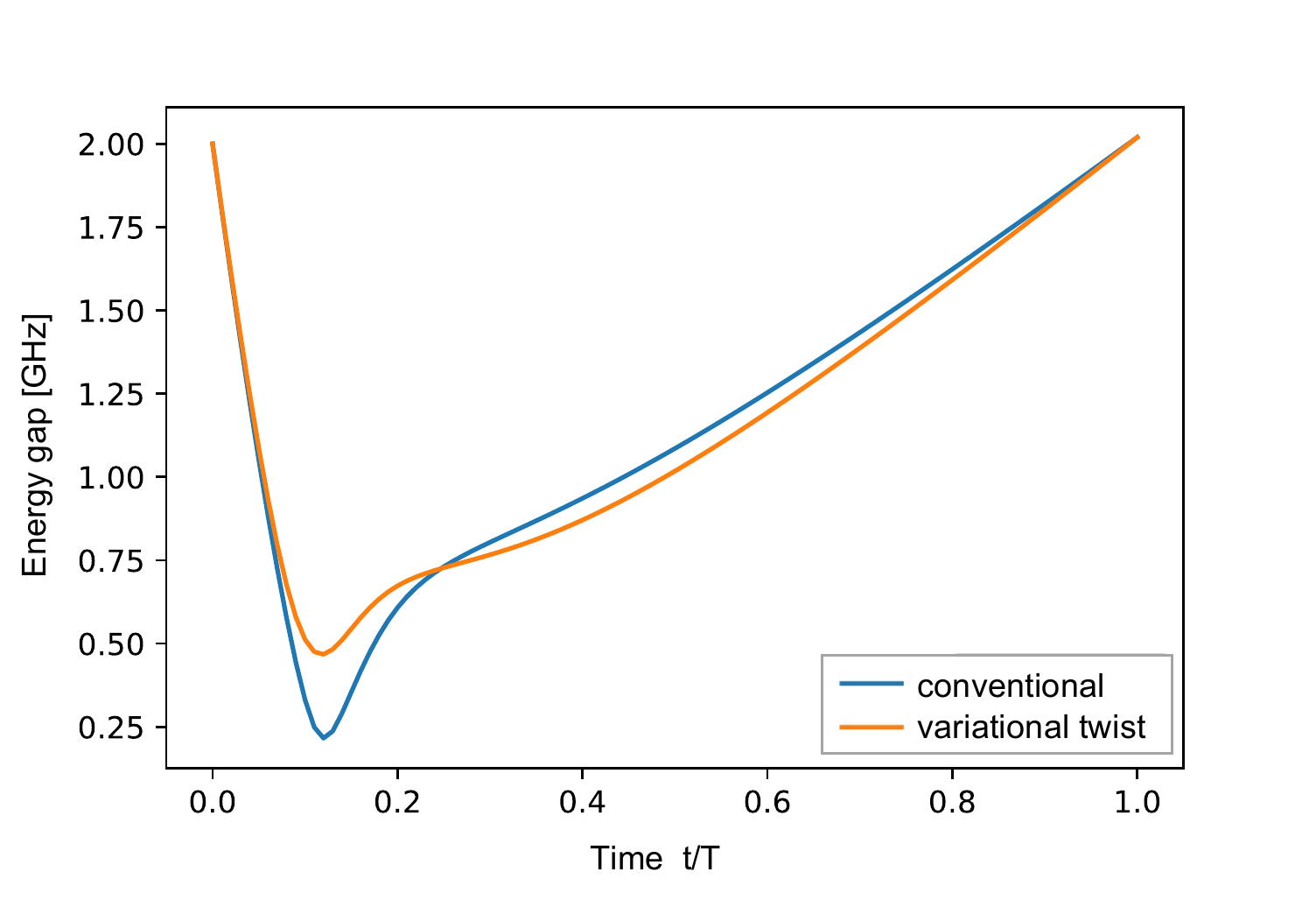}
          (a) Energy gap between the ground state and the first excited state
          \hspace{1.6cm}
        \end{center}
      \end{minipage}

      \begin{minipage}{0.5\hsize}
        \begin{center}
          \includegraphics[clip, width=7.5cm]{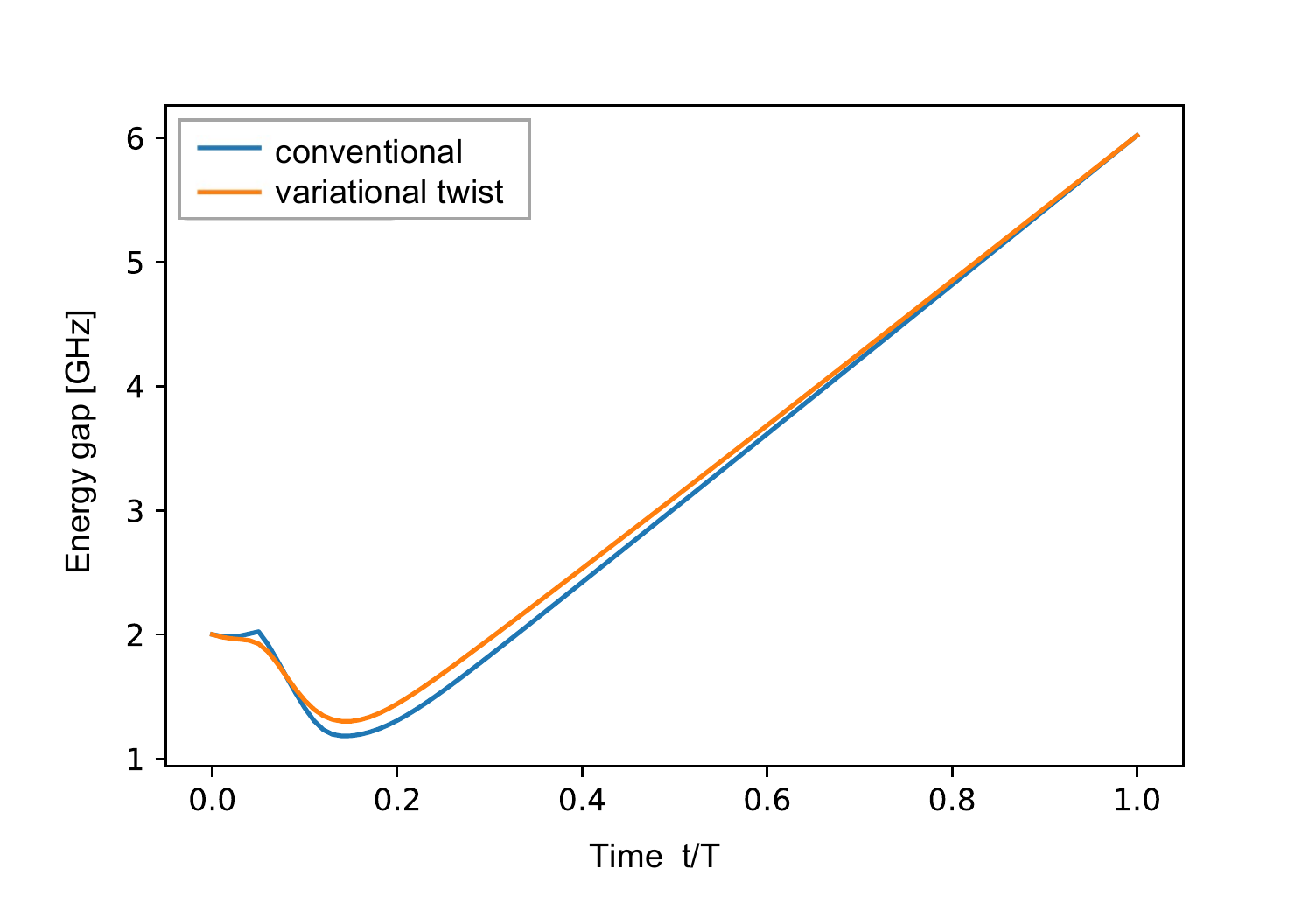}
          (b) Energy gap between the ground state and the second excited state
          \hspace{1.6cm}
        \end{center}
      \end{minipage}
    \end{tabular}
  \end{center}
  \caption{(a) Energy gap between the ground state and the first excited state. (b) Energy gap between the ground state and the second excited state. In (a) and (b), the learning rate $\alpha=0.001$, the decoherence rate $\gamma=10^{-4}$, and the number of steps is $500$.}
  \label{fig:energy_gap-deformed-spin-star}
\end{figure}

We set $\gamma=10^{-4}$, $\alpha=0.001$, $J=15$, and $\omega =\omega _1=1$.
Further, the number of steps is $500$.
We plot the energy spectrum of the annealing Hamiltonian at each time in Figure \ref{fig:deformed_spin_star_energy_spectrum}.
From this graph, the energy spectrum does not seem to have changed significantly; however, we observe that
the energy gap between the ground state and the excited states in our scheme is larger than that in the conventional scheme, as shown in Figure \ref{fig:energy_gap-deformed-spin-star} (a).

\begin{figure}[htbp]
  \begin{center}
    \begin{tabular}{c}

      \begin{minipage}{0.5\hsize}
        \begin{center}
          \includegraphics[clip, width=7.5cm]{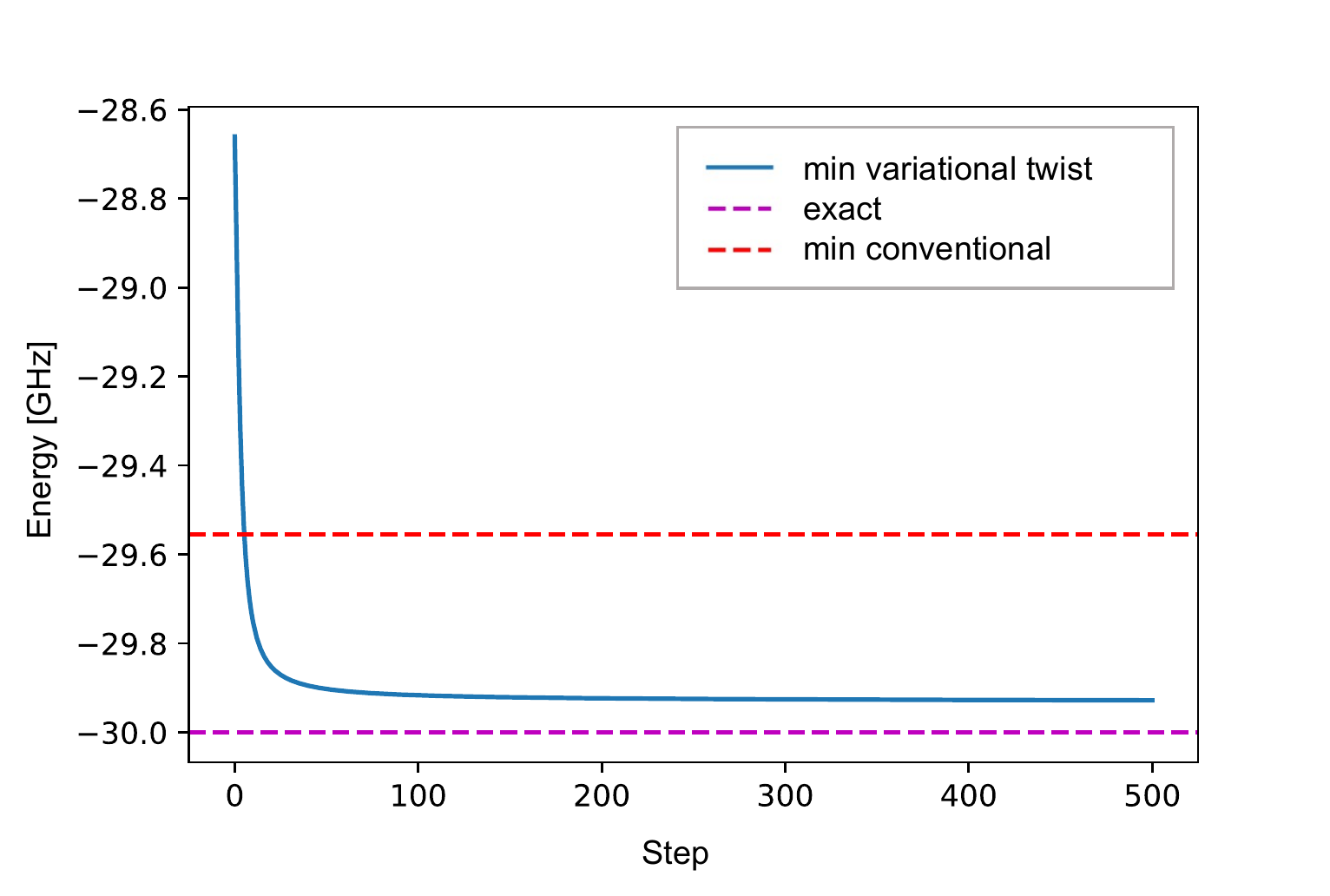}
          (a) Estimation error of the energy against the variational step
          \hspace{1.6cm}
        \end{center}
      \end{minipage}

      \begin{minipage}{0.5\hsize}
        \begin{center}
          \includegraphics[clip, width=7.5cm]{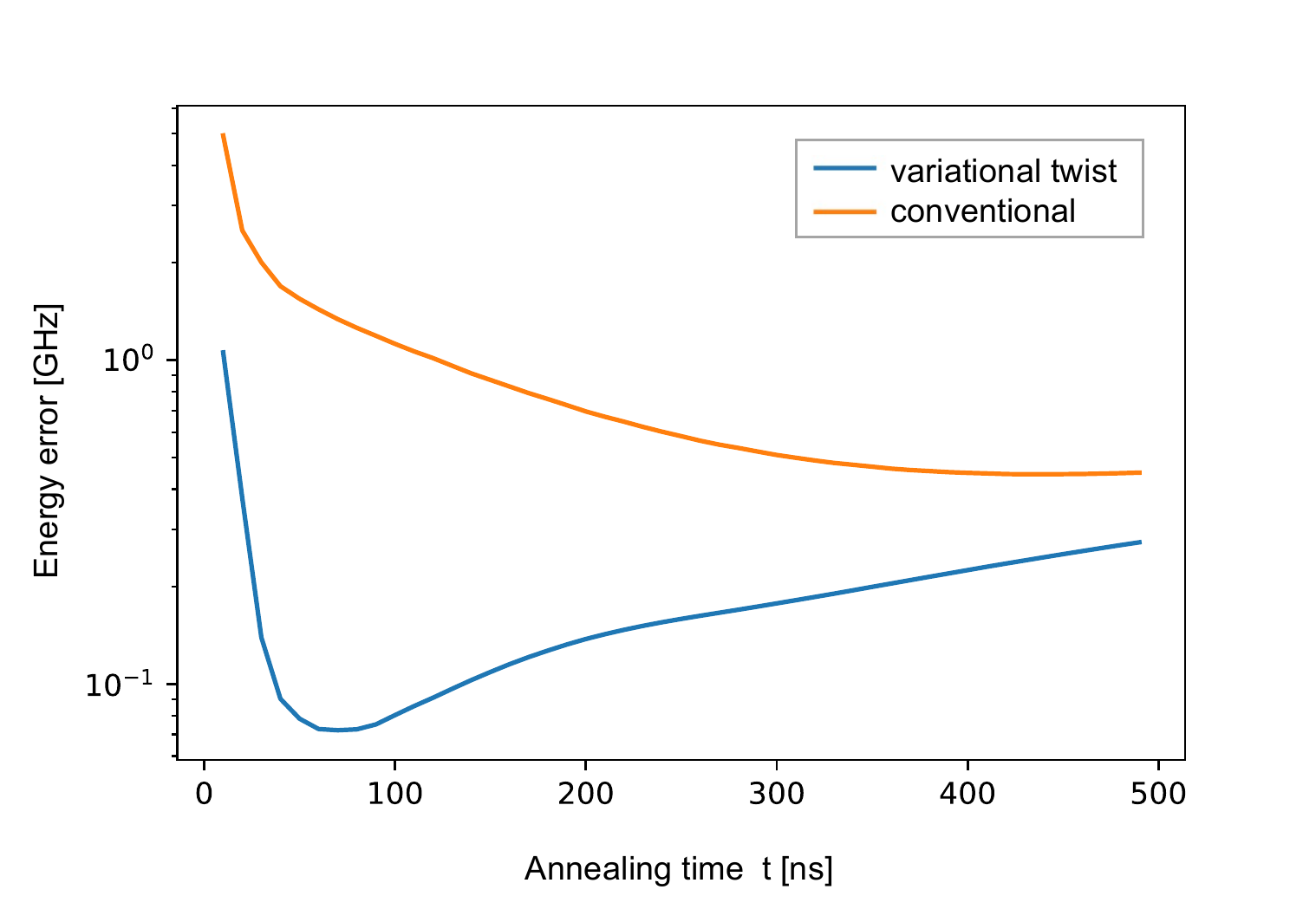}
          (b) Estimation error of the energy against the annealing time $t$
          \hspace{1.6cm}
        \end{center}
      \end{minipage}
    \end{tabular}

  \caption{(a) Estimation error of the energy plotted against the variational step on a log scale. The annealing time is chosen to minimize the energy in our scheme. (b) Estimation error of the energy plotted against the annealing time $t$ on a log scale. In (a) and (b), the learning rate $\alpha=0.001$, the decoherence rate $\gamma=10^{-4}$, and the number of steps is $500$.}
  \label{fig:modify_energy-time_error_modify_energy-step_deformed_spin_star}

  \end{center}

\end{figure}

In Figure \ref{fig:modify_energy-time_error_modify_energy-step_deformed_spin_star}, we plot
the estimation error against the number of variational steps and the annealing time
when we adopt either twisted QA or conventional QA. From these plots, we find that our scheme improves the accuracy by an order of magnitude compared to the conventional scheme.

\begin{figure}[htbp]
  \begin{center}
    \begin{tabular}{c}

      \begin{minipage}{0.5\hsize}
        \begin{center}
          \includegraphics[clip, width=7.5cm]{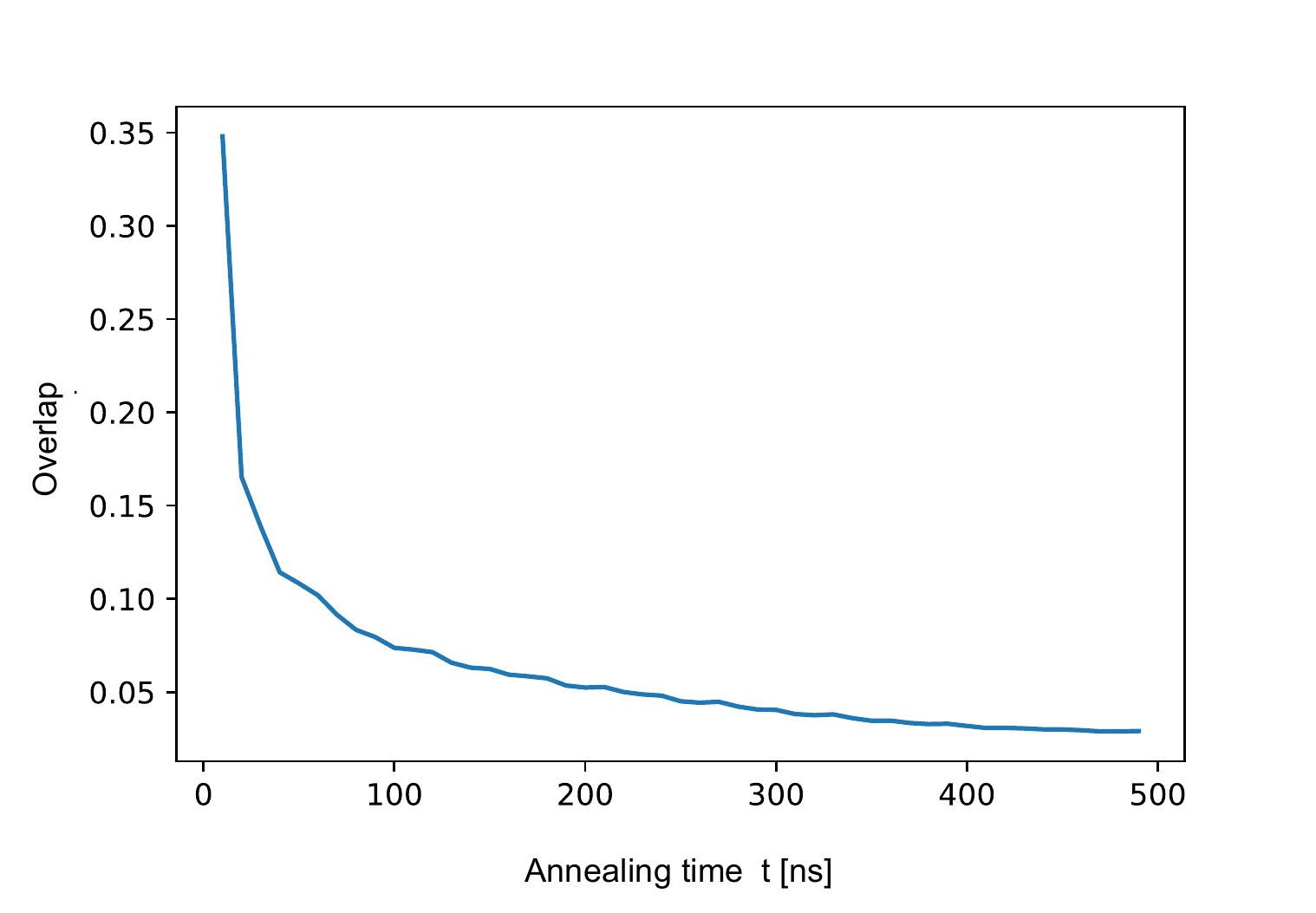}
          (a) Overlap between the initial state of twisted QA and the ground state of the problem Hamiltonian
          \hspace{1.6cm}
        \end{center}
      \end{minipage}

      \begin{minipage}{0.5\hsize}
        \begin{center}
          \includegraphics[clip, width=7.5cm]{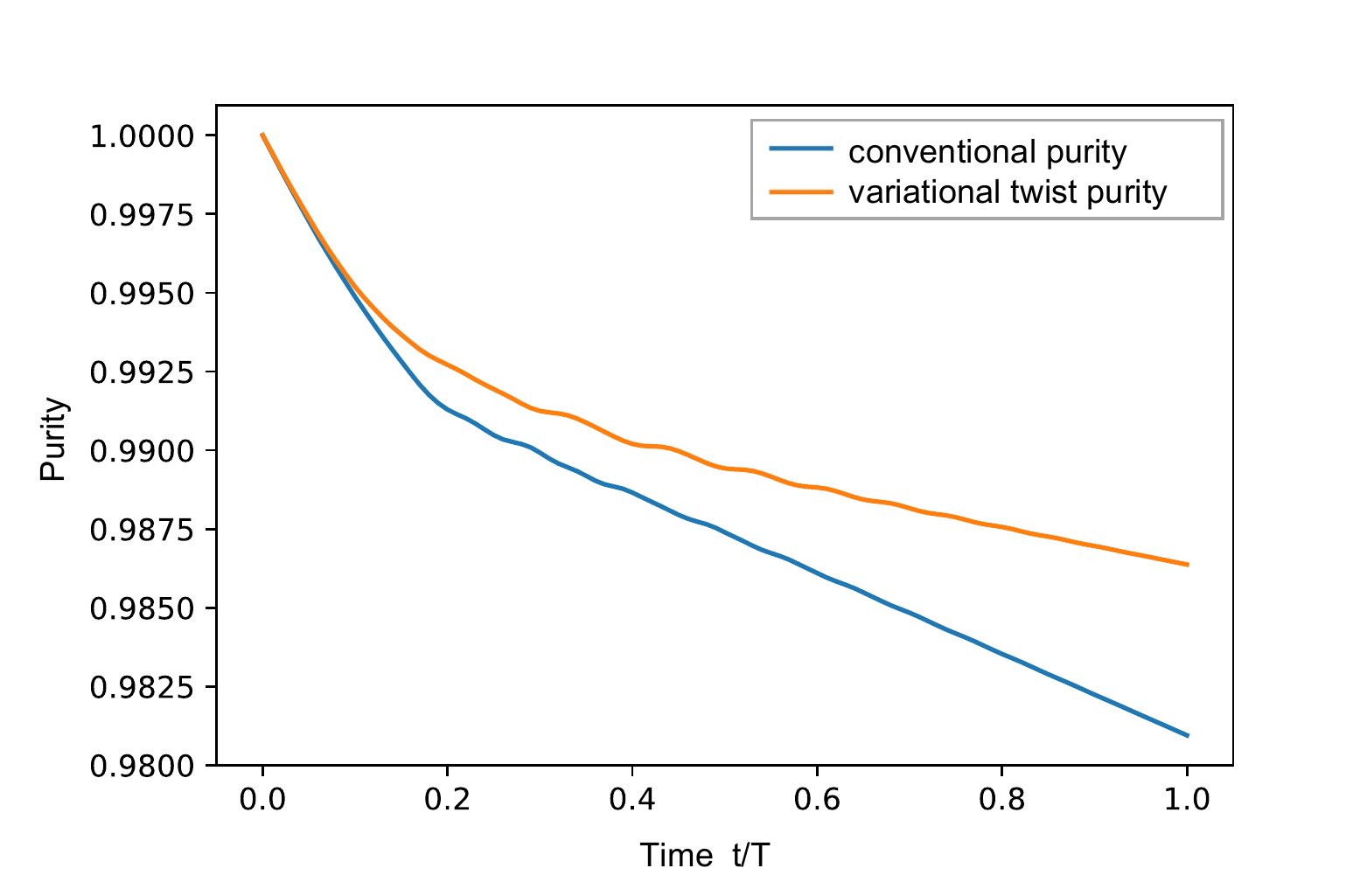}
          (b) Purity against the annealing time
          \hspace{1.6cm}
        \end{center}
      \end{minipage}
    \end{tabular}
  \end{center}
  \caption{(a) Overlap between the initial state of twisted QA and the ground state of the problem Hamiltonian. (b) Purity plotted against the annealing time. The annealing time is chosen to minimize the energy in our scheme. In (a) and (b), the learning rate $\alpha=0.001$, the decoherence rate $\gamma=10^{-4}$, and the number of steps is $500$.}
  \label{fig:overlap_purity_deformed_spin_star}
\end{figure}

In Figure \ref{fig:overlap_purity_deformed_spin_star} (a), we plot the overlap between the initial state of twisted QA and the ground state of the problem Hamiltonian.
From this Figure, we confirm that the overlap with the deformed spin star model is smaller than that with the hydrogen molecule.
We investigate the effects of decoherence and non-adiabatic transitions.
First, we plot the purity to show the effect of decoherence in Figure \ref{fig:overlap_purity_deformed_spin_star} (b).
We can see that the purity of our scheme is slightly higher than that of the conventional scheme. However, as the difference in purity between them is small, we conclude that our scheme cannot significantly suppress the effect of decoherence in this case.

Furthermore, we show the energy gap between the ground state and the first excited state in Figure \ref{fig:energy_gap-deformed-spin-star} (a).
Comparing the smallest energy gaps in this graph, we can see that the difference between our method and the conventional method is around two orders of magnitude. In the case of the energy gap between the ground state and the second excited state, the results obtained for the conventional scheme and our scheme are nearly the same.

\begin{figure}[htbp]
  \begin{center}
    \begin{tabular}{c}

     \begin{minipage}{0.5\hsize}
        \begin{center}
          \includegraphics[clip, width=7.5cm]{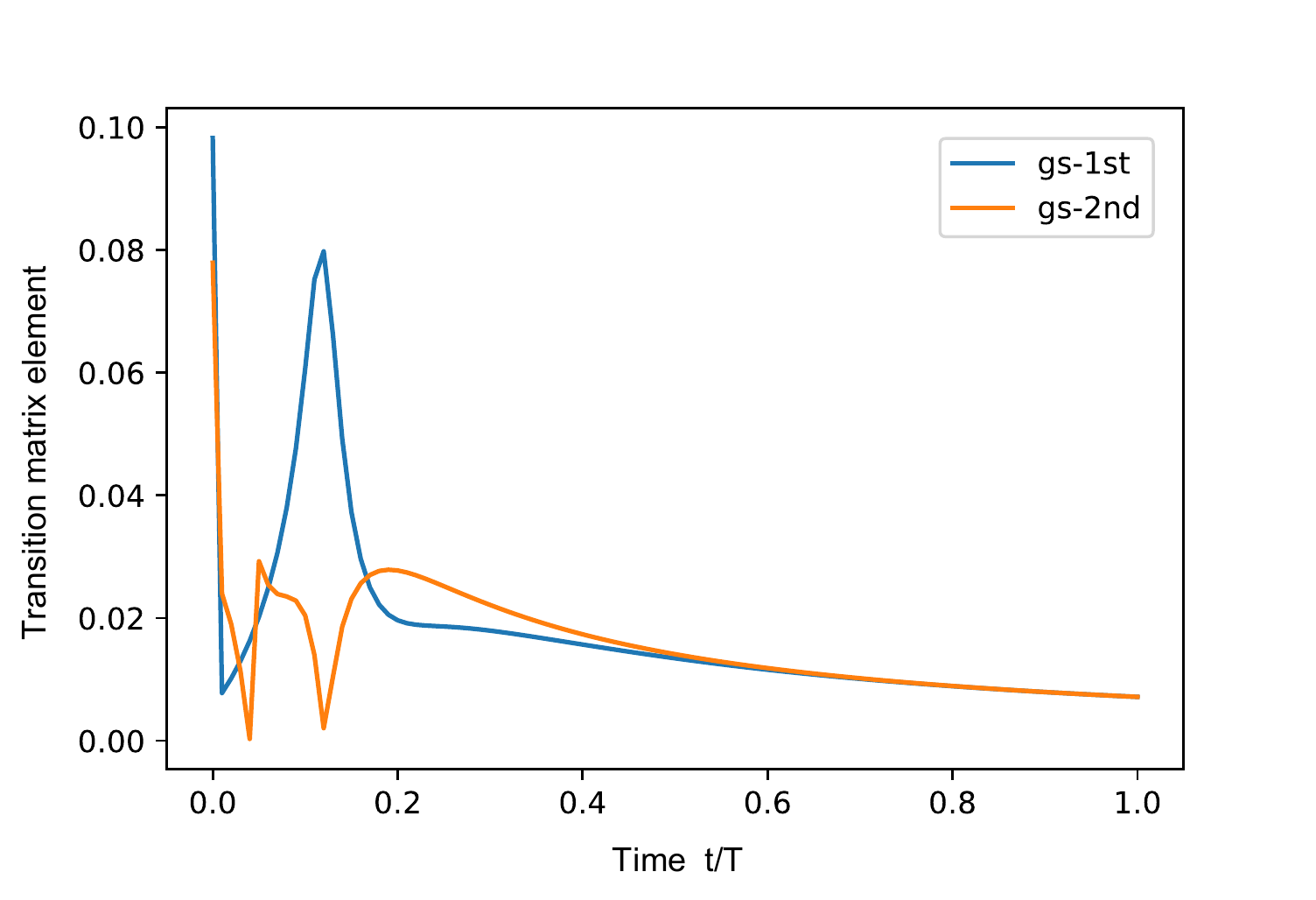}
          (a) Element of the transition matrix of the Hamiltonian in the conventional case
          \hspace{1.6cm}
        \end{center}
      \end{minipage}
      \begin{minipage}{0.5\hsize}
        \begin{center}
          \includegraphics[clip, width=7.5cm]{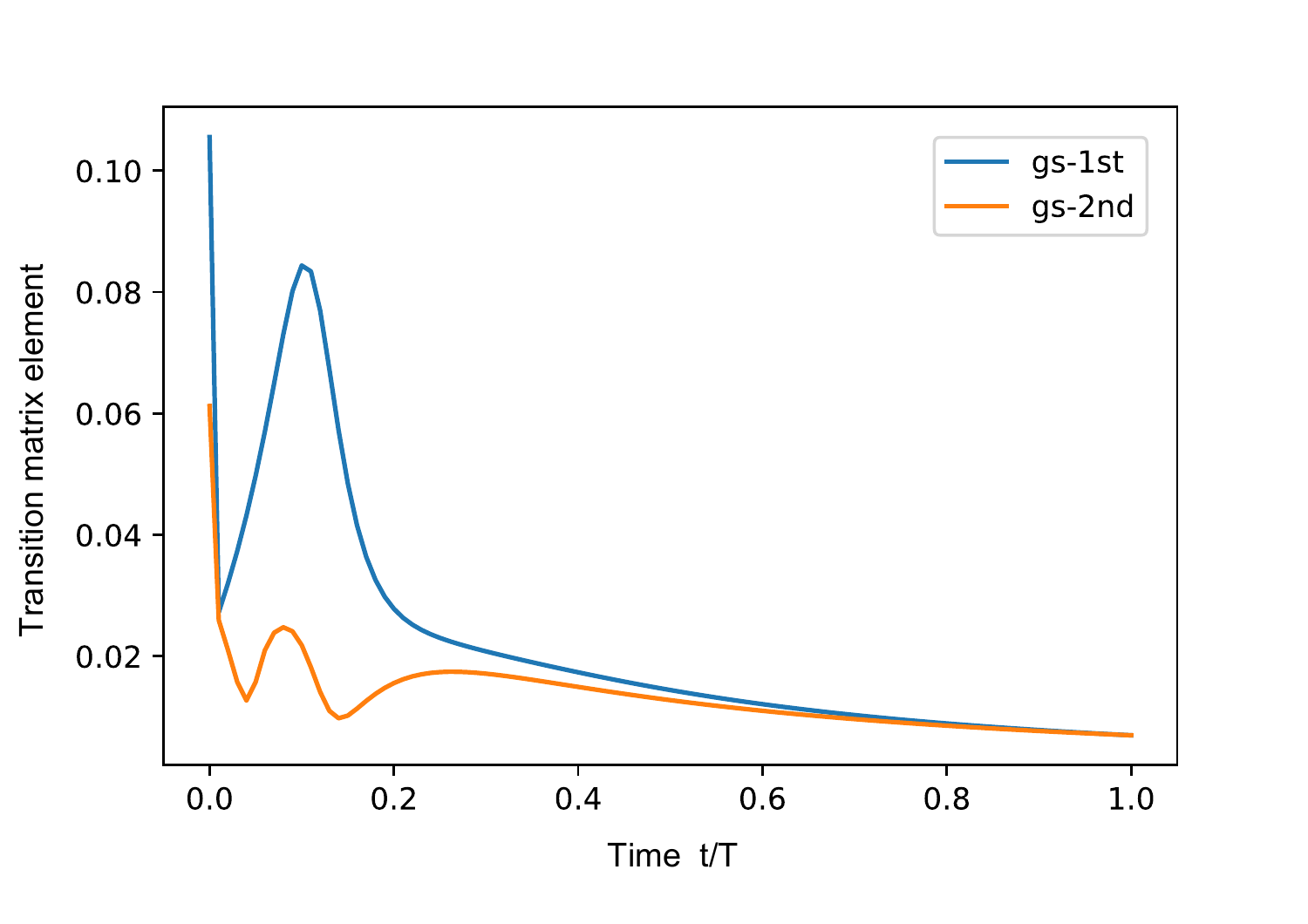}
          (b) Element of the transition matrix of the Hamiltonian in the twisted case
          \hspace{1.6cm}
        \end{center}
      \end{minipage}

    \end{tabular}
  \end{center}
  \caption{(a) Element of the transition matrix of the derivative of the Hamiltonian in the conventional case. (b) Element of the transition matrix of the derivative of the Hamiltonian in the twisted case. In (a) and (b), the learning rate $\alpha=0.001$, the decoherence rate $\gamma=10^{-4}$, and the number of steps is $500$.}
  \label{fig:transition_numerator_defomed_spin_star}
\end{figure}

Finally, we consider
the transition matrix elements of the derivative of the Hamiltonian
from the ground state to the $j$-th excited state.
The transition matrices between the ground state and the first and second excited states
are plotted for conventional QA and our method in Figure \ref{fig:transition_numerator_defomed_spin_star}.
The difference between the conventional method and our method is rather small.

From these results, we conclude that the improvement in the accuracy of our scheme arises from the increase in the energy gap owing to the twisting operations.

\section{Conclusion}\label{sec:conclusion}
In this paper, we proposed a variational method for determining a suitable driving Hamiltonian for QA.
We employed a parameterized twist operator to change the driving Hamiltonian, where each spin was rotated with some angle characterized by the twist parameters.
Starting from the conventional (transverse-field) driving Hamiltonian,
we updated the parameters to minimize the energy of the state after QA until the energy converged to a certain value.
We found that the behavior of the states in our scheme strongly depends on whether the ground state of the problem Hamiltonian is close to a product state or a highly entangled state.
If the ground state of the problem Hamiltonian is close to a product state, our variational method tends to find a driving Hamiltonian whose ground state has a large overlap with the target ground state. By contrast, if the ground state of the problem Hamiltonian is highly entangled, our method tends to find a driving Hamiltonian that has a larger energy gap during QA.
In summary, our QA approach of using variational methods with twist operators can estimate the ground state energy of the problem Hamiltonian with higher accuracy than conventional QA.

\ack

This work was supported by MEXT's Leading Initiative for Excellent Young Researchers and JST PRESTO (Grant No. JPMJPR1919), Japan. This paper is partly
based on the results obtained from a project, JPNP16007,
commissioned by the New Energy and Industrial Technology Development Organization (NEDO), Japan.

\section*{References}

\bibliography{main}

\providecommand{\noopsort}[1]{}\providecommand{\singleletter}[1]{#1}%
\begin{thebibliography}{10}

\bibitem{apolloni1989quantum}
Bruno Apolloni, C~Carvalho, and Diego De~Falco.
\newblock Quantum stochastic optimization.
\newblock {\em Stochastic Processes and their Applications}, 33(2):233--244,
  1989.

\bibitem{finnila1994quantum}
Aleta~Berk Finnila, MA~Gomez, C~Sebenik, Catherine Stenson, and Jimmie~D Doll.
\newblock Quantum annealing: A new method for minimizing multidimensional
  functions.
\newblock {\em Chemical Physics Letters}, 219(5-6):343--348, 1994.

\bibitem{kadowaki1998quantum}
Tadashi Kadowaki and Hidetoshi Nishimori.
\newblock Quantum annealing in the transverse ising model.
\newblock {\em Physical Review E}, 58(5):5355, 1998.

\bibitem{farhi2000quantum}
Edward Farhi, Jeffrey Goldstone, Sam Gutmann, and Michael Sipser.
\newblock Quantum computation by adiabatic evolution.
\newblock {\em arXiv preprint quant-ph/0001106}, 2000.

\bibitem{farhi2001quantum}
Edward Farhi, Jeffrey Goldstone, Sam Gutmann, Joshua Lapan, Andrew Lundgren,
  and Daniel Preda.
\newblock A quantum adiabatic evolution algorithm applied to random instances
  of an np-complete problem.
\newblock {\em Science}, 292(5516):472--475, 2001.

\bibitem{lechner2015quantum}
Wolfgang Lechner, Philipp Hauke, and Peter Zoller.
\newblock A quantum annealing architecture with all-to-all connectivity from
  local interactions.
\newblock {\em Science advances}, 1(9):e1500838, 2015.

\bibitem{kumar2018quantum}
Vaibhaw Kumar, Gideon Bass, Casey Tomlin, and Joseph Dulny.
\newblock Quantum annealing for combinatorial clustering.
\newblock {\em Quantum Information Processing}, 17(2):1--14, 2018.

\bibitem{choi2011minor}
Vicky Choi.
\newblock Minor-embedding in adiabatic quantum computation: Ii. minor-universal
  graph design.
\newblock {\em Quantum Information Processing}, 10(3):343--353, 2011.

\bibitem{bravyi2002fermionic}
Sergey~B Bravyi and Alexei~Yu Kitaev.
\newblock Fermionic quantum computation.
\newblock {\em Annals of Physics}, 298(1):210--226, 2002.

\bibitem{verstraete2005mapping}
Frank Verstraete and J~Ignacio Cirac.
\newblock Mapping local hamiltonians of fermions to local hamiltonians of
  spins.
\newblock {\em Journal of Statistical Mechanics: Theory and Experiment},
  2005(09):P09012, 2005.

\bibitem{seeley2012bravyi}
Jacob~T Seeley, Martin~J Richard, and Peter~J Love.
\newblock The bravyi-kitaev transformation for quantum computation of
  electronic structure.
\newblock {\em The Journal of chemical physics}, 137(22):224109, 2012.

\bibitem{tranter2015b}
Andrew Tranter, Sarah Sofia, Jake Seeley, Michael Kaicher, Jarrod McClean, Ryan
  Babbush, Peter~V Coveney, Florian Mintert, Frank Wilhelm, and Peter~J Love.
\newblock The b ravyi--k itaev transformation: Properties and applications.
\newblock {\em International Journal of Quantum Chemistry}, 115(19):1431--1441,
  2015.

\bibitem{xia2017electronic}
Rongxin Xia, Teng Bian, and Sabre Kais.
\newblock Electronic structure calculations and the ising hamiltonian.
\newblock {\em The Journal of Physical Chemistry B}, 122(13):3384--3395, 2017.

\bibitem{eyring1935activated}
Henry Eyring.
\newblock The activated complex in chemical reactions.
\newblock {\em The Journal of Chemical Physics}, 3(2):107--115, 1935.

\bibitem{kurihara2014quantum}
Kenichi Kurihara, Shu Tanaka, and Seiji Miyashita.
\newblock Quantum annealing for clustering.
\newblock {\em arXiv preprint arXiv:1408.2035}, 2014.

\bibitem{berwald2018computing}
Jesse~J Berwald, Joel~M Gottlieb, and Elizabeth Munch.
\newblock Computing wasserstein distance for persistence diagrams on a quantum
  computer.
\newblock {\em arXiv preprint arXiv:1809.06433}, 2018.

\bibitem{joseph2021two}
David Joseph, Adam Callison, Cong Ling, and Florian Mintert.
\newblock Two quantum ising algorithms for the shortest-vector problem.
\newblock {\em Physical Review A}, 103(3):032433, 2021.

\bibitem{johnson2011quantum}
Mark~W Johnson, Mohammad~HS Amin, Suzanne Gildert, Trevor Lanting, Firas Hamze,
  Neil Dickson, Richard Harris, Andrew~J Berkley, Jan Johansson, Paul Bunyk,
  et~al.
\newblock Quantum annealing with manufactured spins.
\newblock {\em Nature}, 473(7346):194--198, 2011.

\bibitem{kudo2018constrained}
Kazue Kudo.
\newblock Constrained quantum annealing of graph coloring.
\newblock {\em Physical Review A}, 98(2):022301, 2018.

\bibitem{adachi2015application}
Steven~H Adachi and Maxwell~P Henderson.
\newblock Application of quantum annealing to training of deep neural networks.
\newblock {\em arXiv preprint arXiv:1510.06356}, 2015.

\bibitem{hu2019quantum}
Feng Hu, Ban-Nan Wang, Ning Wang, and Chao Wang.
\newblock Quantum machine learning with d-wave quantum computer.
\newblock {\em Quantum Engineering}, 1(2):e12, 2019.

\bibitem{kudo2020localization}
Kazue Kudo.
\newblock Localization in the constrained quantum annealing of graph coloring.
\newblock {\em Journal of the Physical Society of Japan}, 89(6):064001, 2020.

\bibitem{morita2008mathematical}
Satoshi Morita and Hidetoshi Nishimori.
\newblock Mathematical foundation of quantum annealing.
\newblock {\em Journal of Mathematical Physics}, 49(12):125210, 2008.

\bibitem{messiah1961quantum}
Albert Messiah.
\newblock {\em Quantum mechanics}, volume~1.
\newblock John Wiley \& Sons Incorporated, 1961.

\bibitem{messiah1962quantum}
Albert Messiah.
\newblock {\em Quantum mechanics: volume II}.
\newblock North-Holland Publishing Company Amsterdam, 1962.

\bibitem{roland2005noise}
Jeremie Roland and Nicolas~J Cerf.
\newblock Noise resistance of adiabatic quantum computation using random matrix
  theory.
\newblock {\em Physical Review A}, 71(3):032330, 2005.

\bibitem{aaberg2005quantum}
Johan {\AA}berg, David Kult, and Erik Sj{\"o}qvist.
\newblock Quantum adiabatic search with decoherence in the instantaneous energy
  eigenbasis.
\newblock {\em Physical Review A}, 72(4):042317, 2005.

\bibitem{albash2015decoherence}
Tameem Albash and Daniel~A Lidar.
\newblock Decoherence in adiabatic quantum computation.
\newblock {\em Physical Review A}, 91(6):062320, 2015.

\bibitem{childs2001robustness}
Andrew~M Childs, Edward Farhi, and John Preskill.
\newblock Robustness of adiabatic quantum computation.
\newblock {\em Physical Review A}, 65(1):012322, 2001.

\bibitem{sarandy2005adiabatic}
MS~Sarandy and DA~Lidar.
\newblock Adiabatic quantum computation in open systems.
\newblock {\em Physical review letters}, 95(25):250503, 2005.

\bibitem{susa2018exponential}
Yuki Susa, Yu~Yamashiro, Masayuki Yamamoto, and Hidetoshi Nishimori.
\newblock Exponential speedup of quantum annealing by inhomogeneous driving of
  the transverse field.
\newblock {\em Journal of the Physical Society of Japan}, 87(2):023002, 2018.

\bibitem{susa2018quantum}
Yuki Susa, Yu~Yamashiro, Masayuki Yamamoto, Itay Hen, Daniel~A Lidar, and
  Hidetoshi Nishimori.
\newblock Quantum annealing of the p-spin model under inhomogeneous transverse
  field driving.
\newblock {\em Physical Review A}, 98(4):042326, 2018.

\bibitem{seki2012quantum}
Yuya Seki and Hidetoshi Nishimori.
\newblock Quantum annealing with antiferromagnetic fluctuations.
\newblock {\em Physical Review E}, 85(5):051112, 2012.

\bibitem{seki2015quantum}
Yuya Seki and Hidetoshi Nishimori.
\newblock Quantum annealing with antiferromagnetic transverse interactions for
  the hopfield model.
\newblock {\em Journal of Physics A: Mathematical and Theoretical},
  48(33):335301, 2015.

\bibitem{matsuzaki2021direct}
Yuichiro Matsuzaki, Hideaki Hakoshima, Kenji Sugisaki, Yuya Seki, and Shiro
  Kawabata.
\newblock Direct estimation of the energy gap between the ground state and
  excited state with quantum annealing.
\newblock {\em Japanese Journal of Applied Physics}, 2021.

\bibitem{pudenz2014error}
Kristen~L Pudenz, Tameem Albash, and Daniel~A Lidar.
\newblock Error-corrected quantum annealing with hundreds of qubits.
\newblock {\em Nature communications}, 5(1):1--10, 2014.

\bibitem{suzuki2020proposal}
Takayuki Suzuki and Hiromichi Nakazato.
\newblock A proposal of noise suppression for quantum annealing.
\newblock {\em arXiv preprint arXiv:2006.13440}, 2020.

\bibitem{chen2011experimental}
Hongwei Chen, Xi~Kong, Bo~Chong, Gan Qin, Xianyi Zhou, Xinhua Peng, and
  Jiangfeng Du.
\newblock Experimental demonstration of a quantum annealing algorithm for the
  traveling salesman problem in a nuclear-magnetic-resonance quantum simulator.
\newblock {\em Physical Review A}, 83(3):032314, 2011.

\bibitem{nakahara2013lectures}
Mikio Nakahara.
\newblock {\em Lectures on quantum computing, thermodynamics and statistical
  physics}, volume~8.
\newblock World Scientific, 2013.

\bibitem{matsuzaki2020quantum}
Yuichiro Matsuzaki, Hideaki Hakoshima, Yuya Seki, and Shiro Kawabata.
\newblock Quantum annealing with capacitive-shunted flux qubits.
\newblock {\em Japanese Journal of Applied Physics}, 59(SG):SGGI06, 2020.

\bibitem{crosson2014different}
Elizabeth Crosson, Edward Farhi, Cedric Yen-Yu Lin, Han-Hsuan Lin, and Peter
  Shor.
\newblock Different strategies for optimization using the quantum adiabatic
  algorithm.
\newblock {\em arXiv preprint arXiv:1401.7320}, 2014.

\bibitem{goto2020excited}
Hayato Goto and Taro Kanao.
\newblock Excited-state adiabatic quantum computation started with vacuum
  states.
\newblock {\em arXiv preprint arXiv:2005.07511}, 2020.

\bibitem{hormozi2017nonstoquastic}
Layla Hormozi, Ethan~W Brown, Giuseppe Carleo, and Matthias Troyer.
\newblock Nonstoquastic hamiltonians and quantum annealing of an ising spin
  glass.
\newblock {\em Physical review B}, 95(18):184416, 2017.

\bibitem{muthukrishnan2016tunneling}
Siddharth Muthukrishnan, Tameem Albash, and Daniel~A Lidar.
\newblock Tunneling and speedup in quantum optimization for
  permutation-symmetric problems.
\newblock {\em Physical Review X}, 6(3):031010, 2016.

\bibitem{brady2017necessary}
Lucas~T Brady and Wim van Dam.
\newblock Necessary adiabatic run times in quantum optimization.
\newblock {\em Physical Review A}, 95(3):032335, 2017.

\bibitem{somma2012quantum}
Rolando~D Somma, Daniel Nagaj, and M{\'a}ria Kieferov{\'a}.
\newblock Quantum speedup by quantum annealing.
\newblock {\em Physical review letters}, 109(5):050501, 2012.

\bibitem{das2008colloquium}
Arnab Das and Bikas~K Chakrabarti.
\newblock Colloquium: Quantum annealing and analog quantum computation.
\newblock {\em Reviews of Modern Physics}, 80(3):1061, 2008.

\bibitem{susa2021variational}
Yuki Susa and Hidetoshi Nishimori.
\newblock Variational optimization of the quantum annealing schedule for the
  lechner-hauke-zoller scheme.
\newblock {\em Physical Review A}, 103(2):022619, 2021.

\bibitem{matsuura2021variationally}
Shunji Matsuura, Samantha Buck, Valentin Senicourt, and Arman Zaribafiyan.
\newblock Variationally scheduled quantum simulation.
\newblock {\em Physical Review A}, 103(5):052435, 2021.

\bibitem{2109.13043}
Gianluca Passarelli, Rosario Fazio, and Procolo Lucignano.
\newblock Transitionless quantum annealing in a dissipative environment.
\newblock {\em arXiv preprint arXiv:2109.13043}, 2021.

\bibitem{peruzzo2014variational}
Alberto Peruzzo, Jarrod McClean, Peter Shadbolt, Man-Hong Yung, Xiao-Qi Zhou,
  Peter~J Love, Al{\'a}n Aspuru-Guzik, and Jeremy~L O’brien.
\newblock A variational eigenvalue solver on a photonic quantum processor.
\newblock {\em Nature communications}, 5(1):1--7, 2014.

\bibitem{mcclean2016theory}
Jarrod~R McClean, Jonathan Romero, Ryan Babbush, and Al{\'a}n Aspuru-Guzik.
\newblock The theory of variational hybrid quantum-classical algorithms.
\newblock {\em New Journal of Physics}, 18(2):023023, 2016.

\bibitem{johansson2012qutip}
J~Robert Johansson, Paul~D Nation, and Franco Nori.
\newblock Qutip: An open-source python framework for the dynamics of open
  quantum systems.
\newblock {\em Computer Physics Communications}, 183(8):1760--1772, 2012.

\bibitem{johansson184nation}
JR~Johansson.
\newblock Nation pd\& nori f. 2013 qutip 2: a python framework for the dynamics
  of open quantum systems.
\newblock {\em Comp. Phys. Comm}, 184:1234.

\bibitem{puri2017quantum}
Shruti Puri, Christian~Kraglund Andersen, Arne~L Grimsmo, and Alexandre Blais.
\newblock Quantum annealing with all-to-all connected nonlinear oscillators.
\newblock {\em Nature communications}, 8(1):1--9, 2017.

\bibitem{jansen2007bounds}
Sabine Jansen, Mary-Beth Ruskai, and Ruedi Seiler.
\newblock Bounds for the adiabatic approximation with applications to quantum
  computation.
\newblock {\em Journal of Mathematical Physics}, 48(10):102111, 2007.

\bibitem{marcos2010coupling}
D~Marcos, Martijn Wubs, JM~Taylor, R~Aguado, Mikhail~D Lukin, and
  Anders~S{\o}ndberg S{\o}rensen.
\newblock Coupling nitrogen-vacancy centers in diamond to superconducting flux
  qubits.
\newblock {\em Physical review letters}, 105(21):210501, 2010.

\bibitem{twamley2010superconducting}
Jason Twamley and Sean~Duncan Barrett.
\newblock Superconducting cavity bus for single nitrogen-vacancy defect centers
  in diamond.
\newblock {\em Physical Review B}, 81(24):241202, 2010.

\bibitem{zhu2011coherent}
Xiaobo Zhu, Shiro Saito, Alexander Kemp, Kosuke Kakuyanagi, Shin-ichi Karimoto,
  Hayato Nakano, William~J Munro, Yasuhiro Tokura, Mark~S Everitt, Kae Nemoto,
  et~al.
\newblock Coherent coupling of a superconducting flux qubit to an electron spin
  ensemble in diamond.
\newblock {\em Nature}, 478(7368):221--224, 2011.

\bibitem{zhu2014observation}
Xiaobo Zhu, Yuichiro Matsuzaki, Robert Ams{\"u}ss, Kosuke Kakuyanagi, Takaaki
  Shimo-Oka, Norikazu Mizuochi, Kae Nemoto, Kouichi Semba, William~J Munro, and
  Shiro Saito.
\newblock Observation of dark states in a superconductor diamond quantum hybrid
  system.
\newblock {\em Nature communications}, 5(1):1--6, 2014.

\bibitem{matsuzaki2015improving}
Yuichiro Matsuzaki, Xiaobo Zhu, Kosuke Kakuyanagi, Hiraku Toida, Takaaki
  Shimooka, Norikazu Mizuochi, Kae Nemoto, Kouichi Semba, WJ~Munro, Hiroshi
  Yamaguchi, et~al.
\newblock Improving the lifetime of the nitrogen-vacancy-center ensemble
  coupled with a superconducting flux qubit by applying magnetic fields.
\newblock {\em Physical Review A}, 91(4):042329, 2015.

\bibitem{cai2015analysis}
H~Cai, Y~Matsuzaki, K~Kakuyanagi, H~Toida, X~Zhu, N~Mizuochi, K~Nemoto,
  K~Semba, WJ~Munro, S~Saito, et~al.
\newblock Analysis of the spectroscopy of a hybrid system composed of a
  superconducting flux qubit and diamond nv- centers.
\newblock {\em Journal of Physics: Condensed Matter}, 27(34):345702, 2015.

\end{thebibliography}
\end{document}